\documentclass[fleqn,usenatbib]{mnras}

\usepackage{newtxtext,newtxmath}

\usepackage[T1]{fontenc}
\usepackage{ae,aecompl}

\usepackage{graphicx}	
\usepackage{amsmath}	
\usepackage{amssymb}	

\usepackage{arydshln}   
\usepackage{multirow}   

\graphicspath{{plots/}}
\hypersetup{draft}

\newcommand{\Gpc}{,{\rm Gpc}}
\newcommand{\Mpc}{,{\rm Mpc}}
\newcommand{\hMpc}{{\ifmmode{,h^{-1}{\rm Mpc}}\else{$h^{-1}$Mpc}\fi}}
\newcommand{\hkpc}{{\ifmmode{,h^{-1}{\rm kpc}}\else{$h^{-1}$kpc}\fi}}
\newcommand{\hMsun}{{\ifmmode{\,h^{-1}{\rm {M_{\odot}}}}\else{$h^{-1}{\rm{M_{\odot}}}$}\fi}}
\newcommand{\Msun}{\,\rm {M_{\odot}}}
\newcommand{\Mstar}{{\ifmmode{,M_{*}}\else{$M_{*}$}\fi}}
\newcommand{\Mhalo}{{\ifmmode{\,M_{\rm halo}}\else{$M_{\rm halo}$}\fi}}
\newcommand{\ltsima}{$\; \buildrel < \over \sim \;$}
\newcommand{\gtsima}{$\; \buildrel > \over \sim \;$}
\newcommand{\lsim}{\lower.5ex\hbox{\ltsima}}
\newcommand{\gsim}{\lower.5ex\hbox{\gtsima}}

\newcommand{\ahf}{\textsc{AHF}}

\newcommand{\gadgetx}{\textsc{Gadget-X}}
\newcommand{\gadgetmusic}{\textsc{Gadget-MUSIC}}
\newcommand{\galacticus}{\textsc{Galacticus}}
\newcommand{\sag}{\textsc{SAG}}
\newcommand{\sage}{\textsc{SAGE}}
\def\kms{\,{\rm km}\,{\rm s}^{-1}}

\defcitealias{Cui2018}{C18}
\defcitealias{Mostoghiu2019}{M19}

\title[The cluster profiles]{The Three Hundred Project: the stellar and gas profiles}

\author[Li et al.]{\parbox{\textwidth}{
Qingyang Li,$^{1}$\thanks{E-mail: qingyli@sjtu.edu.cn (QL)}
Weiguang Cui,$^{2}$\thanks{E-mail: wcui@roe.ac.uk (WC)}
Xiaohu Yang,$^{1,3}$\thanks{E-mail: xyang@sjtu.edu.cn (XY)}
Elena Rasia,$^4$
Romeel Dave,$^{2,5}$ 
Marco De Petris,$^{6,7}$
Alexander Knebe,$^{8,9,10}$ 
John A. Peacock,$^{2}$
Frazer Pearce,$^{11}$ and
Gustavo Yepes$^{8,9}$
}
\vspace{0.4cm}
\\
\parbox{\textwidth}{
$^1$Department of Astronomy, School of
  Physics and Astronomy, Shanghai Jiao Tong University, Shanghai 200240, China\\
$^2$Institute for Astronomy, University of Edinburgh, Royal Observatory, Edinburgh EH9 3HJ, United Kingdom\\
$^3$Tsung-Dao Lee Institute, and Shanghai Key Laboratory for Particle Physics and Cosmology, Shanghai Jiao Tong University, Shanghai 200240, China\\
$^{4}$INAF - Osservatorio Astronomico di Trieste, via Tiepolo 11, I-34143 Trieste, Italy\\
$^5$University of the Western Cape, Bellville, Cape Town 7535, South Africa\\
$^6$Dipartimento di Fisica, Sapienza Universit\`{a} di Roma, p.le Aldo Moro 5, I-00185 Rome, Italy\\
$^7$INFN - Sezione di Roma, P.le A. Moro 2, I-00185 Roma, Italy
$^8$Departamento de F\'isica Te\'{o}rica, M\'{o}dulo 8, Facultad de Ciencias, Universidad Aut\'{o}noma de Madrid, 28049 Madrid, Spain\\
$^9$Centro de Investigaci\'{o}n Avanzada en F\'{\i}sica Fundamental (CIAFF), Universidad Aut\'{o}noma de Madrid, 28049 Madrid, Spain \\
$^10$International Centre for Radio Astronomy Research, The University of Western Australia, 35 Stirling Highway, Crawley, Western Australia 6009, Australia\\
$^{11}$School of Physics \& Astronomy, University of Nottingham, Nottingham NG7 2RD, UK\\
}}
\date{Accepted XXX. Received YYY; in original form ZZZ}

\pubyear{2019}

\begin{document}
\label{firstpage}
\pagerange{\pageref{firstpage}--\pageref{lastpage}}
\maketitle

\begin{abstract}
Using the catalogues of galaxy clusters from The Three Hundred project, modelled with both hydrodynamic simulations, (\gadgetx\ and \gadgetmusic), and semi-analytic models (SAMs), we study the scatter and self-similarity of the profiles and distributions of the baryonic components of the clusters: the stellar and gas mass, metallicity, the stellar age, gas temperature, and the (specific) star formation rate. Through comparisons with observational results, we find that the shape and the scatter of the gas density profiles matches well the observed trends including the reduced scatter at large radii which is a signature of self-similarity suggested in previous studies. One of our simulated sets, \gadgetx, reproduces well the shape of the observed temperature profile, while \gadgetmusic\ has a higher and flatter profile in the cluster centre and a lower and steeper profile at large radii. The gas metallicity profiles from both simulation sets, despite following the observed trend, have a relatively lower normalisation. The cumulative stellar density profiles from SAMs are in better agreement with the observed result than both hydrodynamic simulations which show relatively higher profiles. The scatter in these physical profiles, especially in the cluster centre region, shows a dependence on the cluster dynamical state and on the cool-core/non-cool-core dichotomy. The stellar age, metallicity and (s)SFR show very large scatter, which are then presented in 2D maps. We also do not find any clear radial dependence of these properties. However, the brightest central galaxies have distinguishable features compared to the properties of the satellite galaxies.
\end{abstract}



\section{Introduction}

Galaxy clusters are the largest gravitationally bound objects in the universe, containing numerous galaxies, intra-cluster medium (ICM) and dark matter. Although the baryonic matter only occupies a small fraction (about the cosmic baryonic fraction $\Omega_b/\Omega_m$) of the total cluster mass, observing galaxies and the ICM at different wavelengths, such as optical, X-ray, radio, allows us to measure several cluster properties and to depict a full picture of the cluster. Moreover, the physical property distributions of gas and stars reflect the effect of different physical processes and the formation of the clusters. Therefore, it is essential to understand their distributions and connections, as well as their dependence on the cluster properties/formation history.

It is well known that the density profile of dark matter haloes is self-similar and can be well described by the NFW \citep{Navarro1997} or Einasto \citep{Merritt2006} fitting formulae. Recent developments in hydrodynamic simulations with baryon models indicate that baryons also play a role in shaping the distribution of dark matter \citep[see][ for a review]{Cui2017book}. Using the hydro-simulated clusters from The Three Hundred project\footnote{\url{https://the300-project.org}} \citep[][hereafter C18]{Cui2018}, \citet[hereafter M19]{Mostoghiu2019} showed that the self-similarity of the total cluster density profile, which starts from $z$ = 2.5, seems to be independent of baryon models \citep[see also][for a similar result but with DM-only simulations]{LeBrun2018}. However, the total density profile shows dependence on the halo formation time/cluster dynamical state. An open question is whether the gas/stars and their physical property profiles, such as temperature, metallicity, also follow a similar trend and these profiles are model-dependent or not.

Clusters have been studied through multi-wavelength observations, such as X-ray and optical. X-ray telescopes can detect the high energy photons scattered by the hot electrons in the ICM via bremsstrahlung emission. These observations provide an insight into the distribution of hot gas \citep[for example][]{Bohringer2010}. With better X-ray telescopes such as XMM-Newton, Chandra and Suzaku, the gas properties are investigated with much greater detail \citep[e.g.][]{Majerowicz2002, Vikhlinin2005, Sato2007}. Similar to the total density profile, it has been suggested that the hot gas beyond the cooling core region in massive clusters also shows a self-similar evolution in the mean profile up to $z \sim 1.9$ \citep[e.g.][]{Vikhlinin2006,McDonald2017}. The gas density profile has been well studied in the outskirts from both numerical and observational studies \citep{Roncarelli2006, Vikhlinin2006, Lemze2008}. These works found that the outer radii profile can be simply fitted by a power-law while in the innermost regions the gas density profile is more cored than the dark matter density profile, even though the trend of the core depends on the cluster dynamical status. Indeed, the cool-core (CC) clusters which have significantly lower temperature gas in the centre and short cooling times show different central gas densities compared with non-cool-core (NCC) clusters. 

Also the temperature and the metallicity profiles generally show self-similar profiles at large radii, which can be fitted by a universal fitting function \citep{Mohr1999,Vikhlinin2005,Baldi2012,Biffi2018a,Ghirardini2019}. The gas temperature profile slowly increases from the outer regions toward the cluster centre \citep[e.g.][]{Vikhlinin2005,Rasmussen2007,Pratt2007,Reiprich2009}. In the cluster central region, the temperature of CC and NCC clusters show distinct trends: the CC cluster drops quickly, while the NCC cluster becomes flat \citep[e.g.][]{Sanderson2006,Finoguenov2007,Dunn2008}. Similarly, the metallicity profile peaked in the centre,  decreases with radius, and becomes flat beyond around $\sim 0.3 \times R_{500}$\footnote{The subscript 500 or 200 used in this paper refers to enclosed overdensities of 500 or 200 times the critical density of the universe} \citep[see][for example]{Tholken2016,Ezer2017,Urban2017,Vogelsberger2018,Lovisari2019}. However, this general trend is more pronounced in CC clusters, which have a significant peak, with respect to NCC objects, which might have a relatively flat profile \citep[e.g.][]{Baldi2007,Leccardi2008}. \citet{Lovisari2019} claimed that the metallicity profile is non-uniform by separating the clusters into dynamically relaxed (high concentration and low centroid-shift) and disturbed (low concentration and high centroid-shift) objects. The relaxed systems show a higher metallicity in the centre compared to disturbed systems.

Optical observations have achieved great success in revealing the distribution of galaxies. However, the faint intracluster light (ICL) is still hard to distinguish due to the sensitivity of current telescopes. Normally, stellar population properties are gauged based on modelling the spectral energy distributions of the galaxies. It is also interesting to see whether the stellar properties present a self-similar profile or not.
Recently, the profile of the star formation rate (SFR) or the specific star formation rate (sSFR) has been investigated in galaxy clusters. \citet{Lagana2018} \citep[see also][for a similar result for higher redshift clusters]{Alberts2016} showed that the SFR seems not to correlate with the projected radius at 0.4 < $z$ < 0.8, while the sSFR may also not follow a growing trend with radius as suggested by \citet{Brodwin2013}, for example.

Therefore, following \citetalias{Mostoghiu2019}, we detail the modelled and observed physical profiles and also focus on the difference between models and observations in this paper. We investigate the profiles of stellar properties (stellar mass, age, metallicity) and also study the thermo- and chemo-dynamical properties of the intracluster medium. We use the multi-modelled clusters from both hydrodynamic simulations and semi-analytic models (SAMs) at z = 0 of The Three Hundred project \citep{Cui2018,Wang2018,Mostoghiu2019,Arthur2019,Ansarifard2020,Haggar2020}. A few comparisons are carried out with observations in the X-ray band and optical, using SDSS data.

This paper is organised as follows. In section 2, we concisely introduce the adopted hydrodynamic simulations and SAMs, as well as the cluster dataset. In section 3, we present the clusters selected from the SDSS 7 catalogue. In section 4, we present the clusters physical profiles and contrast with observed data. Finally, we summarise our conclusions in section 5.

\section{The hydrodynamic simulations and semi-analytic models}
\label{sec:models}

Hydrodynamic simulations and SAMs are described in \citetalias{Cui2018} \citep[see also][for the SAM catalogues]{Knebe2018}, we refer interested readers to those papers for more information. Here, we only briefly summarise some basic details. The 324 regions are centred on galaxy clusters which are initially selected from the MultiDark simulation \citep{Klypin2016}\footnote{The MultiDark simulations are publicly available at \url{https://www.cosmosim.org} database.} -- the dark-matter-only MDPL2 with the cosmological parameters from the Planck mission \citep{Planck2016}. MDPL2 is a periodic cube of comoving size equal to $1.48 \Gpc$ containing $3840^3$ dark matter particles. The selected 324 galaxy clusters are the most massive objects identified at $z=0$ in the parent simulation.
Each re-simulated region has an approximate radius of $\sim 22 \Mpc$ at $z = 0$ which includes the high resolution particles. The outer layer with multiple levels of mass refinement has been generated using the parallel \textsc{Ginnungagap}\footnote{\url{https://github.com/ginnungagapgroup/ginnungagap}} code. The hydrodynamic simulations are run with these initial conditions, while the SAMs galaxies of each re-simulation region are cut out from the MultiDark-Galaxies catalogue \citep{Knebe2018}.

We only use the datasets from two simulation codes -- \gadgetx\ \citep{Murante2010,Rasia2015} and \gadgetmusic\ \citep{MUSICI}.
Both simulation codes are based on the gravity solver of the {\sc GADGET3} Tree-PM code (an updated version of the {\sc GADGET2} code; \citealt{Springel2005}) with smoothed-particle hydrodynamics (SPH) to follow the evolution of the gas component. \gadgetmusic\ uses the classic entropy-conserving SPH formulation with a 40 neighbour spline kernel, while \gadgetx\ includes an improved SPH scheme \citep{Beck2016} with artificial thermal diffusion, time-dependent artificial viscosity, high-order Wendland C4 interpolating kernel and wake-up scheme. \gadgetx\ is also different from \gadgetmusic\ for the treatment of the baryonic components. Stellar evolution and metal enrichment in \gadgetx\ \citep[see][for the original formulation]{Tornatore2007} consider mass-dependent lifetimes of stars \citep{Padovani1993}, the production and evolution of 15 different elements coming from SNIa, SNII and AGB stars with metallicity-dependent radiative cooling \citep{Wiersma2009}. Although both simulations adopt the stellar feedback model from \cite{Springel2003}, \gadgetmusic\ uses a higher wind velocity ($400\kms$) than \gadgetx\ ($350\kms$) for the kinetic stellar feedback. In addition, it also included another mode of thermal feedback -- the evaporation of cold clouds due to SN feedback. While \gadgetx\ models the black hole (BH) growth and implements active galactic nuclei (AGN) feedback \citep{Steinborn2015} unlike \gadgetmusic. We note here that \gadgetmusic\ uses the Salpeter initial mass function \citep[IMF,][]{Salpeter1955} while \gadgetx\ applies the Chabrier IMF \citep{Chabrier2003}.

The aforementioned MDPL2 dark-matter-only simulation has been populated with galaxies \citep{Knebe2018} by three distinct SAMs, i.e. \galacticus\ \citep{Benson2012}, \sag\ \citep{Cora2018}, and \sage\ \citep{Croton2016}. The same 324 regions (using the same radius cut) have also been extracted from the SAMs' haloes and galaxy catalogue that covers the entire simulation volume of the parent MDPL2 simulation. This data set constitutes the counterpart sample from the hydrodynamic catalogue, to which it can be directly compared. All SAMs adopt the Chabrier IMF \citep{Chabrier2003}.

The haloes in each re-simulation region are identified by the Amiga Halo Finder, \ahf\citep{Knollmann2009} using an overdensity threshold of 200 times the critical density of the Universe. In our analysis, we only use the mass-complete clusters. Further, we also recalculate $M_{500}$ and $R_{500}$ for each of the selected clusters using the method presented in \citet{Cui2014}. Additionally, we recalculate both $R_{500}$ and $R_{200}$ for the MDPL2 haloes used in SAM and corresponding to the hydrodynamical simulated sample (see \citetalias{Cui2018} for the matching procedure). The SAM galaxies within the corresponding radii are used for comparisons. Finally, to account for our limited mass resolution we select only objects with stellar mass above 5 $\times\ 10^{10}\,\Msun$ in both modelled and observed samples.

\section{The SDSS galaxy clusters}

We compared with observations of galaxy clusters from the SDSS 7 catalogue. The SDSS 7 catalogue is taken from \citet{Shi_2018} which is based on the \citet{Yang2012} catalogue.
This group catalogue is constructed using the adaptive halo-based group finder and halo masses -- $M_{200}$ -- are assigned to each group using the ranking of either their total characteristic luminosity or total characteristic stellar mass.
It uses the same cosmological parameters as the hydrodynamic simulations and SAMs. To consistently make comparisons, we first apply the same halo mass threshold $M_{200} > 9.47 \times 10^{14}\,\Msun$ to the SDSS group catalog, as the mass-complete simulated sample. In this way, we select out 100 galaxy clusters including 2905 galaxies from the SDSS 7 catalogue. However, 394 galaxies from the \citet{Yang2018} catalogue do not belong to the SDSS catalogue, so they are excluded. In addition, another 22 galaxies are removed from the catalogue because they are not classified as galaxies in \citet{Comparat2017}. 
We furthermore select galaxies with stellar mass $M_\star > 5 \times 10^{10}\,\Msun$ which is consistent with the simulated catalogue. 8 additional clusters which do not have any galaxies above this mass limit are removed. This results in 1142 galaxies.
As the simulation and SAM clusters use the maximum density peak as the centre of the cluster, we consider the brightest/most massive central galaxy (BCG) of the SDSS clusters as the group centre. Finally, the 92 galaxy clusters have 906 satellite galaxies within the projected $r_{200}$. We note here that the mass weighted mean redshift of these galaxies is $z \approx 0.15$.

The stellar population properties of the SDSS galaxies -- age, metallicity, stellar mass and the star formation history -- are given by \citet{Comparat2017}, who performed full spectral fitting on individual spectra making use of 3 different high spectral resolution stellar population models: STELIB \citep{LeBorgne2003}, MILES \citep{Snchez-Blazquez2006,Falcon-Barroso2011, Beifiori2011} and ELODIE \citep{Prugniel2007}. They provided the galaxy properties from different choices of stellar IMF and input stellar libraries, from which we choose the MILES stellar libraries with the Chabrier IMF \citep{Chabrier2003}. We refer to \citet{Comparat2017} for the details of different stellar libraries and IMFs. The SFR and sSFR of these galaxies are taken from \citet{Brinchmann2004}.

Even with this stellar mass cut, the galaxies in clusters at high redshift are still incomplete due to the limiting magnitude of current telescopes, i.e. galaxies of the same mass detected at a lower redshift may not be observed at a higher redshift. We use the following method to complete the galaxy members above a certain mass cut. The basic idea is that if a galaxy was observed in a local group, it can also be observed in a high-redshift group in which the same mass galaxy is beyond the observational limit. We first correct the galaxy's redshift with K and E corrections. We further define $z_{\rm max}$ -- the maximum redshift -- at which a galaxy with a given stellar mass can be observed by the telescope. In this way, every galaxy has its own $z_{\rm max}$. Each member galaxy of a selected cluster is referred as the original galaxy. We select all the clusters of which redshifts are lower than an original galaxy's $z_{\rm max}$. Then, we consider that all these clusters should include this galaxy. However, because there is only one galaxy among these clusters, we conclude that the probability that this original galaxy can be observed in these selected clusters is $1/N_{\rm c}$, where $N_{\rm c}$ is the total number of the selected clusters with $z<z_{\rm max}$. Meanwhile, all the clusters whose redshifts are lower than this galaxy $z_{\rm max}$ should contain this galaxy with the same probability, 1/$N_{\rm c}$. Only satellite galaxies are considered here. These quantities are added to the clusters with redshift larger than the galaxy $z_{\rm max}$ at the same radius of the original galaxy. With this method applied, an additional $\sim$ 1990 galaxies are included in our SDSS sample, which results in 2896 satellite galaxies in total. But we only apply this method to the complete catalogue sample in calculating the stellar number density and mass density. This possibility is directly taken account as the satellite galaxy number for the calculation of satellite number density. While it is multiplied with the original galaxy stellar mass for the calculation of stellar mass density. For the other stellar properties, such as age and metallicity, we only use these member galaxies and do not take this incompleteness correction into account.

\section{results}\label{sec:results}

We present here the scatter and the universality of the physical profiles of galaxy clusters. Physical profiles, separated into stellar and gas components, are presented in Subsections~\ref{subsec:stellar} and \ref{subsec:gas}, respectively. For the stellar physical profiles, we focus on stellar density, age and metallicity, which are all derived only from the satellite galaxies\footnote{Both the BCG and the ICL are not taken into account, unless specified}. For the gaseous physical profiles, we investigate the gas density, temperature, metallicity, SFR and sSFR properties, which are based only on the gas content from the two hydrodynamical runs. It is worth noting that the radius for all the stellar and gas profiles is normalised to $r_{200}$ and $R_{500}$ respectively in order to compare with the observational results. Throughout the paper, $r$ indicates the projected radius (only member galaxies within $R_{200}$ are included in the projection), while $R$ is the distance in 3D. We only select the $x-y$ plane to project these simulated clusters. These profiles are generally shown via medians with the error bar indicating the $16^{th}$ and $84^{th}$ percentiles in each radius bin. The solar metallicity is taken from \citet{Asplund2009} with the value $Z_{\odot}=0.0134$.

\subsection{Stellar physical profiles}\label{subsec:stellar}

\begin{figure}
    \centering
    \includegraphics[width=0.5\textwidth]{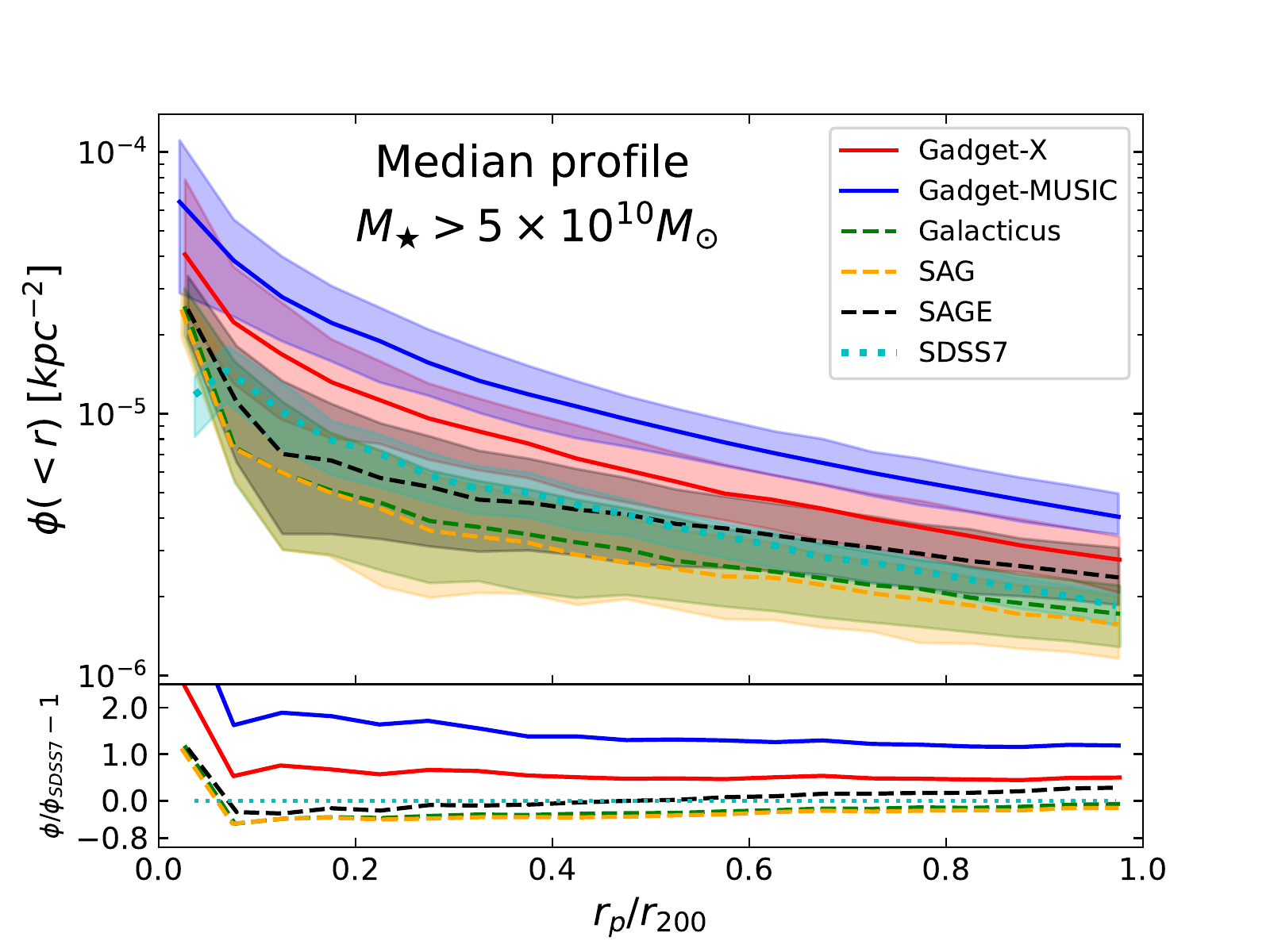}
    \caption{Cumulative galaxy number density as a function of the normalised radius. The same galaxy stellar mass limit (5 $\times\ 10^{10}\,\Msun$) is applied to all modelled and observed galaxies. As indicated in the legend, different colour and line styles represent the median profile from different models and the observed SDSS result. The shaded regions are the $16^{th}$ and $84^{th}$ percentiles of all cluster profiles. The bottom panel shows the residuals compared with the SDSS7 data.}
    \label{fig:stellar_n}
\end{figure}

Due to the fact that all the stellar properties investigated here are derived from satellite galaxies, we first check our data consistency by presenting the cumulative galaxy number density -- $\phi$ -- in the selected galaxy clusters in Fig.~\ref{fig:stellar_n}. Here $\phi$ is defined as $N(<r)/\pi r^2$ where $N(<r)$ is the total galaxy number within $r$. We also show the residuals using each dataset in comparison with the SDSS measurements. Apparently, SAMs are in better agreement with observation than the two hydro-simulations. \gadgetmusic, which does not have AGN feedback, has the highest galaxy number which is $\sim 2$ times higher than the SDSS observation. \gadgetx, even with AGN feedback, still has about 50 per cent more galaxies than the SDSS result. This decreasing trend of the number density profiles indicates that the satellite galaxy number density drops from inner to outer radii. We note here that this galaxy mass cut may bias our results towards high mass galaxies, such as more red or old galaxies which may not be well modelled in the two hydro-simulations (more details can be found in section \ref{secsub:sfr}). The drop of the innermost data point from SDSS clusters could be caused by a projection effect with miss-identification of galaxies (this data point is very close to the edge of the BCG) or by a bias which is coming from the offset between the BCG and the number density peak of the galaxy clusters.

\subsubsection{Stellar mass profile}

\begin{figure}
    \centering
    \includegraphics[width=0.5\textwidth]{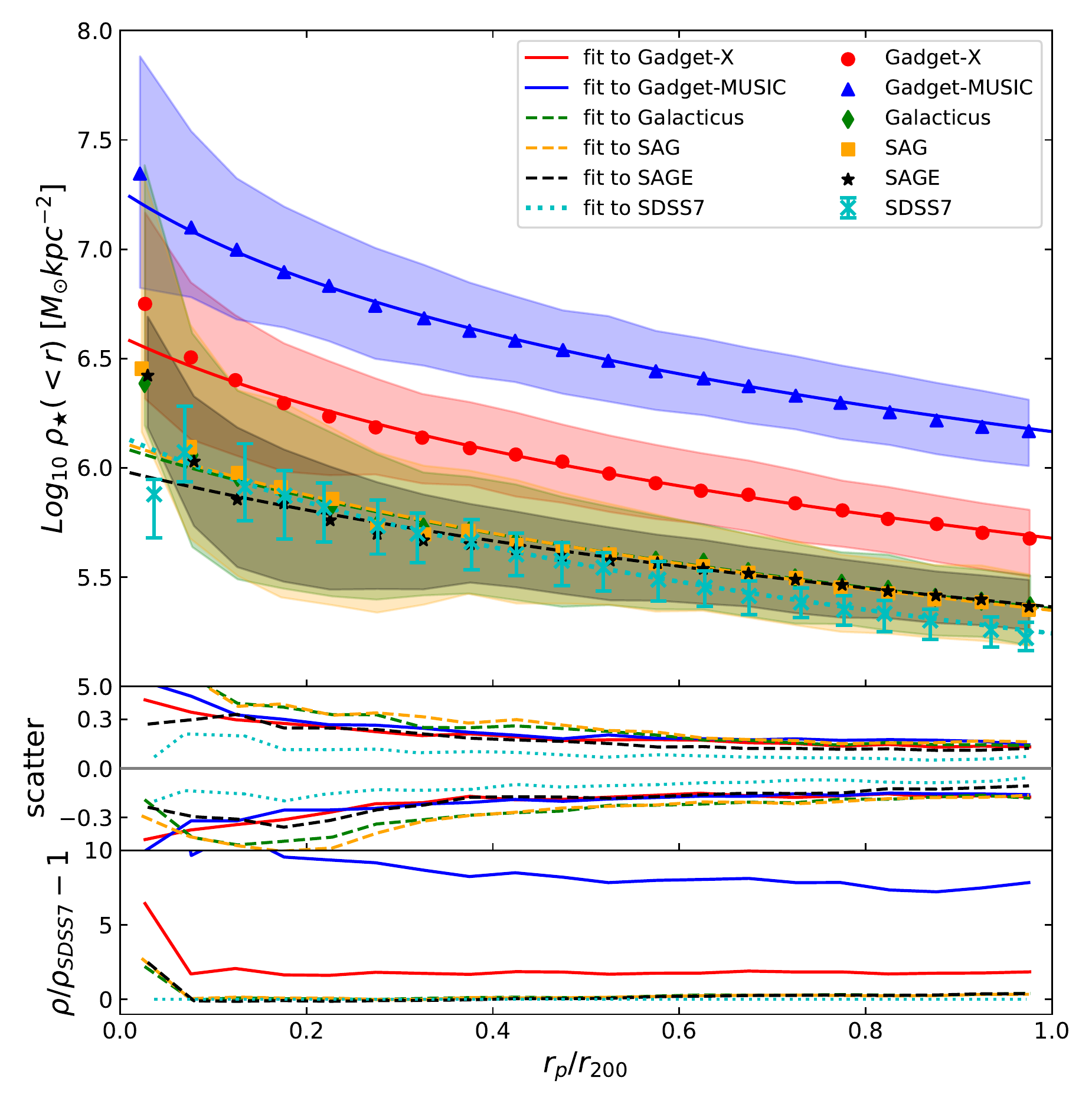}
    \caption{Cumulative stellar mass density profile as a function of the normalised radius. Different models and observed median profiles are shown in different colour symbols, while shaded regions (for models) and  error bars (for the SDSS observation) present $16^{th}$ and $84^{th}$ percentiles of all cluster profiles. The middle panel highlights these scatters. The bottom panel shows the fitting residuals compared with the SDSS result. Fitting results of median data with the formula from \citet{Lokas2001} are presented in Table~\ref{tab:density4_table}.}
    \label{fig:stellar_density4}
\end{figure}

\begin{table*}
	\centering
	\caption{The fitting function \citep[Eq.~\ref{eq:5}, based on][]{Lokas2001} and parameters for Fig.~\ref{fig:stellar_density4}. $M_{*, 200}$ is set to the total stellar mass of satellite galaxies within $r_{200}$. Here, we use the median value of $r_{200}$ and $M_{*, 200}$ for these clusters (shown in the second and third rows) in the fitting function and fit the free parameter c with the median density profiles of each model.}
	\label{tab:density4_table}
	\begin{tabular}{c c c c c c c}
		\hline
		parameters & \gadgetx\ & \gadgetmusic\ & \galacticus\ & \sag\ & \sage\ & SDSS\\
		\hline
		$r_{200} [10^3\ \rm kpc]$  & $2.246$ & $2.251$ & $2.251$ & $2.251$ & $2.251$ & $2.294$\\
		$M_{*,200}\ [10^{12} \Msun]$ & $7.558$ & $23.319$ & $3.615$ & $3.546$ & $3.681$ & $2.893$\\
		\hline
		free parameter c & $3.486$ & $4.896$ & $2.368$ & $2.539$ & $1.810$ & $3.356$ \\
 		\hline
	\end{tabular}
\end{table*}

The stellar mass density profile indicates how the stars/galaxies are distributed in the cluster environments. As the galaxy cluster is at the final stage of structure formation, this stellar mass density profile could potentially be a powerful tool to constrain galaxy formation and evolution. However, it is not easy to measure the observed stellar component in the cluster accurately. This is because the intra-cluster light is very faint, normally below the telescope detection limit, and it can contribute a significant amount of stellar mass \citep[see][for example]{Cui2014a}, especially at the cluster centre region. Therefore, we only use the satellite galaxy stellar mass from both models and observation here to estimate this stellar mass density profile. 

The cumulative stellar mass density profile, which uses the galaxy mass above the mass cut, is presented in Fig.~\ref{fig:stellar_density4}. The fitting function with the form of the NFW profile \citep{Navarro1997} is obtained from \citet{Lokas2001}:
\begin{equation}
    \centering
    M(s) = M_{200}\ g(c)\left(\ln{(1+cs)}-\frac{cs}{1+cs}\right)
    \label{eq:1}
\end{equation}
where
    \begin{equation}
        s = \frac{r}{r_{200}},
    \end{equation}
    \begin{equation}
        c = \frac{r_{200}}{r_s},
    \end{equation}
    \begin{equation}
        g(c) = \frac{1}{\ln{(1+c)} - \frac{c}{1 + c}}.
    \end{equation}
Finally, the density profile can be expressed as: 
\begin{equation}
    \rho = \frac{M(s)}{V},
    \label{eq:5}
\end{equation}
where volume $V$ equals $\pi r^2$ in projection and $4/3\pi R^3$ in real space. We note here that this fitting function was originally used for dark matter density profiles. It seems that the stellar density profile follows a similar profile to dark matter. So we use this function as the fitting function, only replacing the total mass, $M_{200}$, by the total satellite stellar mass, $M_{*,200}$. We exclude the innermost datapoint from these fits. The scatter and fitting residuals are respectively presented in the middle and bottom panel of Fig.~\ref{fig:stellar_density4}. The residuals are calculated by comparing with the fitted SDSS result. In agreement with Fig.~\ref{fig:stellar_n}, the stellar mass density profiles are also basically in alignment with the SDSS result, except \gadgetmusic\ and \gadgetx. The almost constant shift of the \gadgetmusic\ and \gadgetx\ profiles with respect to the result from SDSS indicates that (1) the AGN feedback has a homogeneous effect on the galaxy mass that does not depend on the distance to the cluster centre; (2) the density profile from \gadgetx, even with AGN feedback, is still about 2 times higher than the profile from SDSS; (3) \gadgetmusic, which does not include AGN feedback and has a weaker SN feedback \citep[as implied by its higher satellite galaxy stellar mass function in][]{Cui2018} compared to \gadgetx, presents a much higher (about 4 times) stellar density profile.
The error bars are at a level of $\sim$0.5 dex depending on the models. This indicates that, like the halo density profile, the stellar density profile of the galaxy cluster is almost universal. We study the origin of this scatter by separating our clusters into relaxed and un-relaxed clusters in Appendix \ref{app:1}, which only has a weak impact for the outer radii. The fitting parameters are listed in Table~\ref{tab:density4_table}.

\subsubsection{Stellar age distribution}

\begin{figure*}
    \centering
    \includegraphics[width=1.\textwidth]{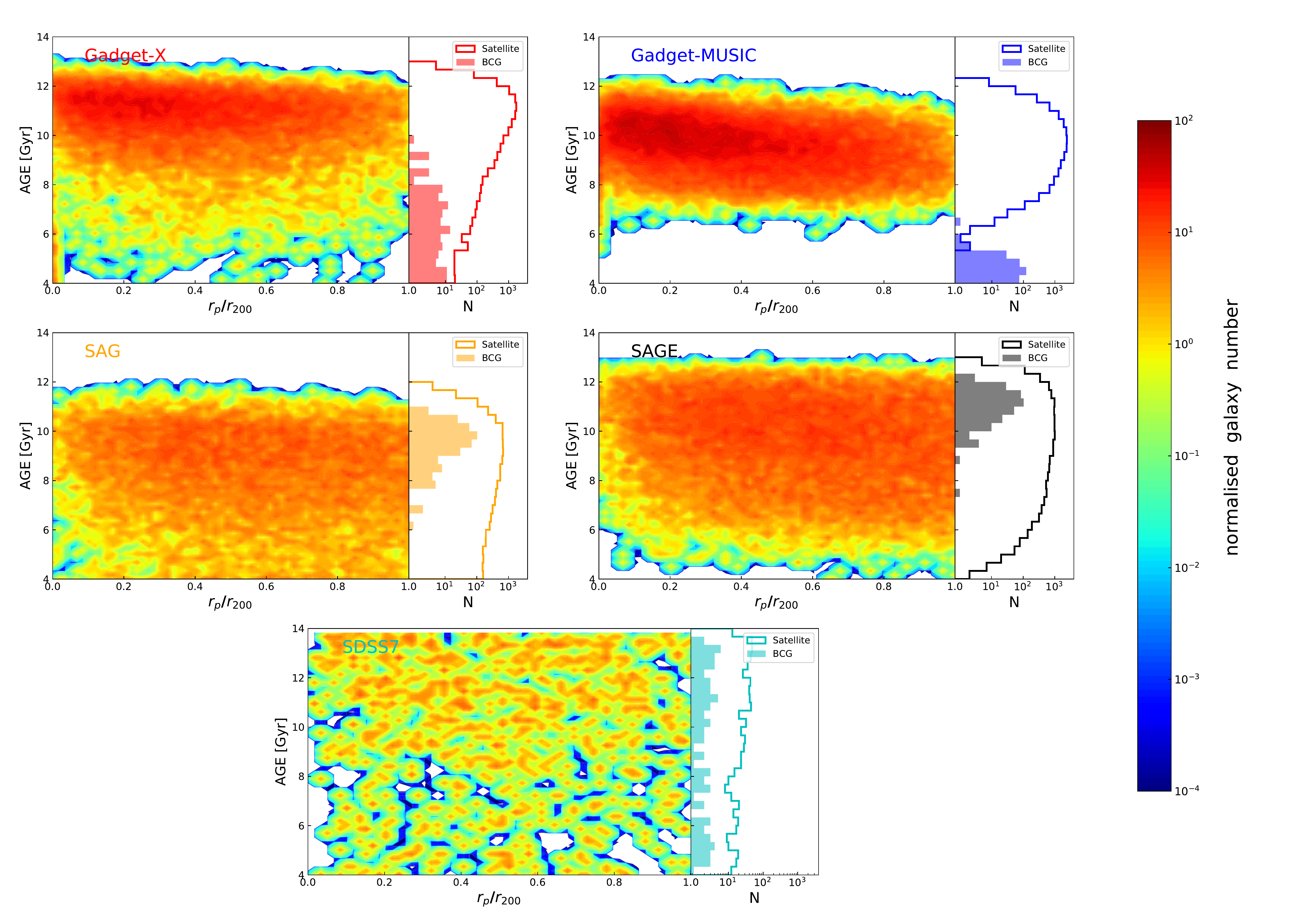}
    \caption{The distribution of satellite galaxy age as a function of radius inside the clusters. No BCGs are included in the left-hand panels. The right-hand sub-panels show the age histograms for the BCGs (filled) and satellite galaxies (solid). The colour bar shows the normalised galaxy number density with respect to the total cluster number. A Gaussian filter with standard deviation of 0.5 pixels is applied to smooth these maps.}
    \label{fig:age2D}
\end{figure*}

For modelled galaxies, age is the mass-weighted age of all stars inside the $z=0$ galaxy. We find that the mean and mass-weighted ages are very similar as the stellar particles have similar masses. The age of an observed galaxy is derived from model fitting of the galaxy spectrum \citep[see details in][]{Comparat2017}. For SDSS galaxies with low S/N spectra, SED model fits will tend to overestimate the age. This is because the stellar model fit is independent of cosmology. Therefore, we exclude about 120 galaxies which have their age older than the Universe (13.79 Gyr) in this analysis. The satellite galaxies' age distributions is shown in Fig.~\ref{fig:age2D}. \galacticus\ is excluded from this plot due to the lack of galaxy age. Histograms of the galaxy ages are shown in the right-hand sub-panels. The age of BCGs from hydrodynamic simulations is defined as the mass-weighted star particle age within $0.015\times R_{200}$. We investigated varying this to the larger radius of $0.05 \times R_{200}$ and did not find any significant changes in the BCG ages.

Firstly, there is a very large scatter in the age distributions of both the SDSS and modelled galaxies. This indicates a significant mix of young and old galaxies at all radii in the cluster environment. Although we exclude galaxies older than the Universe, there is a noticeable fraction of very old galaxies. This is due to the model fitting. Secondly, the satellite galaxies in both hydro-simulations are primarily dominated by old galaxies with ages of $\sim$ 11 Gyr, which are not apparent in either the SAMs or SDSS. Thirdly, unlike the SAMs and the SDSS, the BCGs from both hydro-simulations are much younger than their satellite galaxies. The reason could be that star formation is not fully quenched by AGN activity in the cluster centre, which is especially clear in the \gadgetmusic\ run. 


\subsubsection{Stellar metallicity distribution}

\begin{figure*}
    \centering
    \includegraphics[width=1.\textwidth]{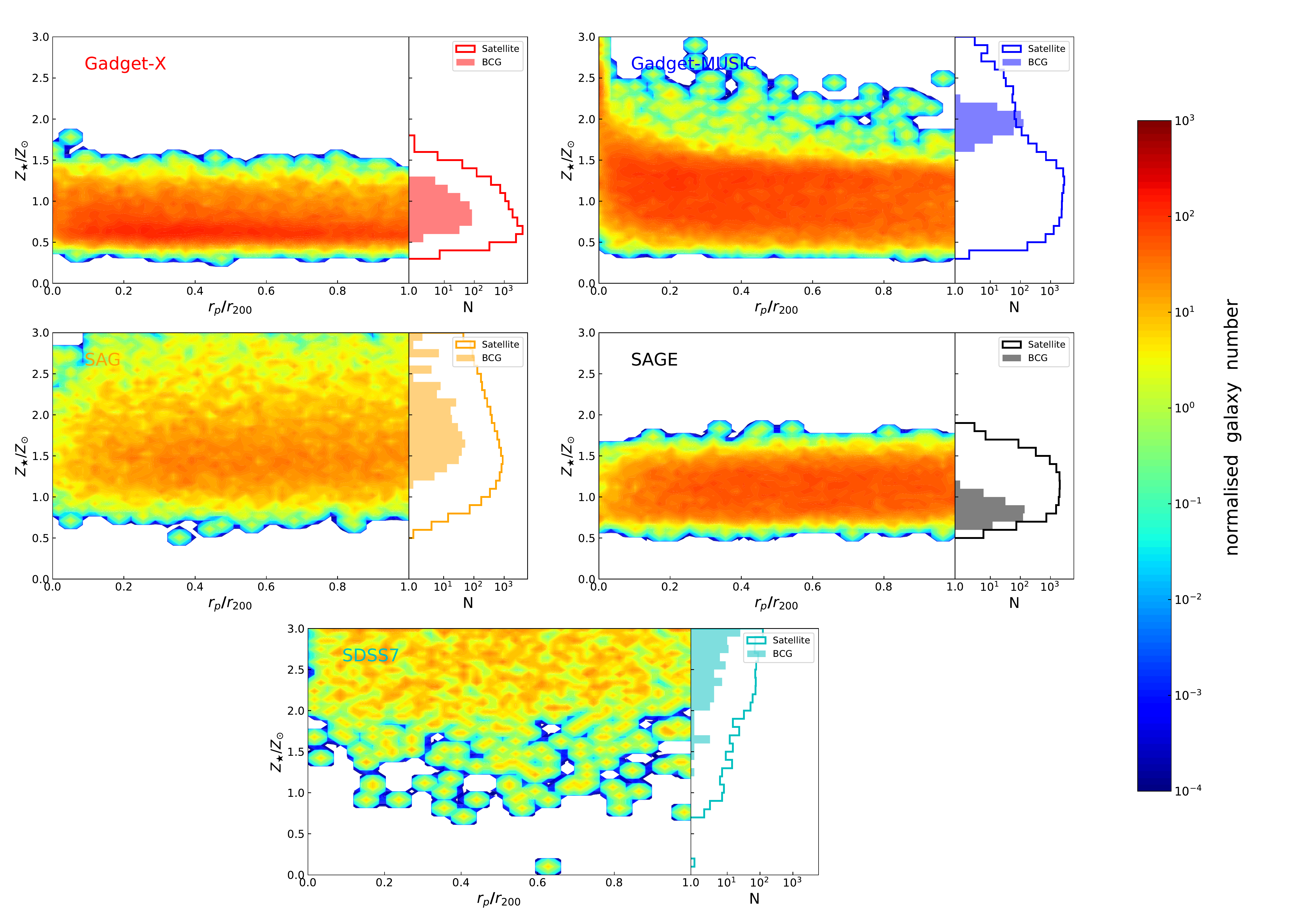}
    \caption{Similar to the age distribution in Fig.~\ref{fig:age2D} but for the stellar metallicity distribution of the satellite galaxies inside the clusters. The BCG distributions are only included in the right-hand panels as filled steps.}
    \label{fig:starz2D}
\end{figure*}

Stellar metal enrichment is mainly determined by stellar nucleosynthesis, which is correlated with the chosen IMF, initial metallicity and age. Therefore, the galaxy stellar metallicity is directly related to its age.

The galaxy stellar metallicity distributions with respect to the solar metallicity for the two hydrodynamic simulations, SAMs and the SDSS 7 galaxies are presented in Fig.~\ref{fig:starz2D}. The metallicity of the BCGs in the hydrodynamic simulations is defined as the mass-weighted star particles' metallicity within 0.015$R_{\rm 200}$. Again, we do not find any significant change on the BCG metallicities when this limiting radius is taken to be ten times larger. The striking point is that the disagreement between the models and observation is quite large. The galaxy metallicity from models is generally less than the solar metallicity, while the majority of SDSS galaxies have metallicity $\sim 1.5 - 2 \times Z_\odot$. The sharp cut-off in the SDSS metallicity map is due to the limitation of the SED fitting model. When we compare the galaxy metallicity distribution with their age distribution, we find that the modelled galaxies with a greater age tend to have a lower metallicity. This can be understood as the metallicity is dominated by younger stars, which have a higher initial metallicity because of their later formation. However, SDSS galaxies tend to have both older age and higher metallicity, which does not fit into this picture. This could be due to the intrinsic simple stellar population (SSP) fitting, which may provide different views on the galaxy age and metallicity. In particular, this SSP fitting is strongly model dependent \citep{Comparat2017}. The ELODIE-type models give a distribution of metallicities stretching towards sub-solar values while the MILES-based models used in this work remain more concentrated at solar metallicities. The STELIB library grants a smaller coverage in metallicity, hence model results are confined between half-solar and twice-solar in chemical composition. The range in ages found using STELIB-based models is larger and extends to younger ages with respect to the other two models. If we put aside the uncertainty in the SSP fitting and assume that the disagreement between the models and observation is real, one possible solution for models is to form more young stars. However this will also bring down the age profile. Therefore, this dilemma should perhaps instead be solved via other methods such as a higher metal production in SN feedback. 
 
\subsection{Gas profiles} \label{subsec:gas}

\begin{figure}
    \centering
    \includegraphics[width=0.5\textwidth]{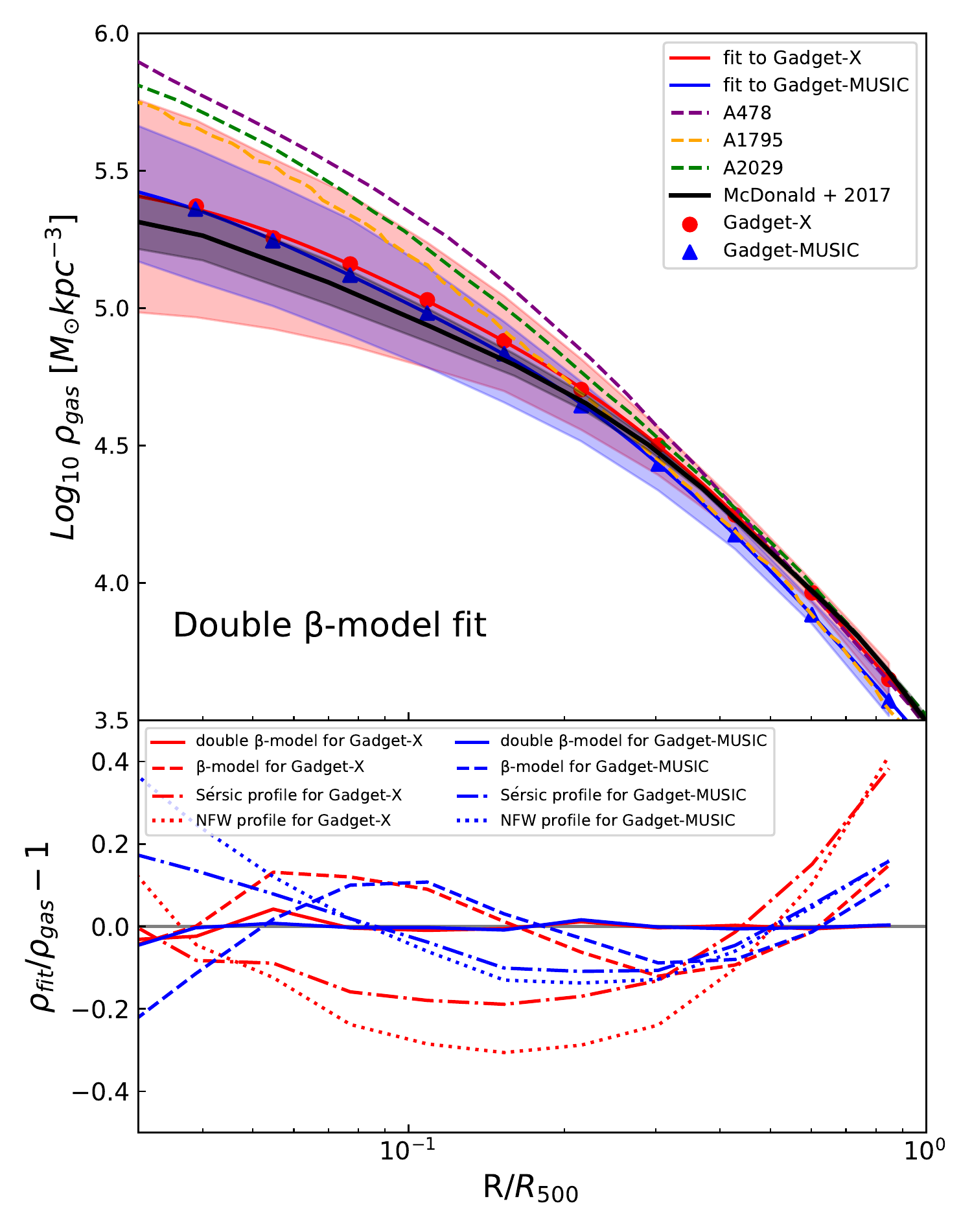}
    \caption{Radial gas mass density profile for two hydrodynamic simulations. Red circles and blue triangles respectively represent the median profiles from \gadgetx\ and the \gadgetmusic\ runs with solid lines for their best fit with a double $\beta$-model. Shaded areas for two hydrodynamic simulations show the $16^{th}$ and $84^{th}$ percentiles of all cluster profiles. Purple, orange and green dashed lines show gas density profiles of three galaxy clusters: A478, A1795 and A2029 respectively, which are taken from \citet{Vikhlinin2006}. The black solid line shows the median gas density profile with 1 sigma uncertainty for galaxy clusters at 0.0 < z < 0.1 from \citet{McDonald2017}. Bottom panels are the residuals between different the fitting function and the median data points for \gadgetx\ (red) and \gadgetmusic\ (blue). Solid, dashed, dash-dotted and dotted lines respectively show different fitting functions: double $\beta$-model, $\beta$-model, S$\rm{\Acute{e}}$rsic profile and NFW profile.}
    \label{fig:gas_density0}
\end{figure}

\begin{table*}
	\centering
	\caption{The double $\beta$-model fitting function \citep[based on][]{Mohr1999} and fitting parameters for Fig. \ref{fig:gas_density0}.}
	\label{tab:density0_table}
	\begin{tabular}{c c c c c c}
	    \hline
        fitting parameters & $\rho_1[\Msun /kpc^3]$ & $R_1[\rm R_{500}]$ & $\rho_2[\Msun /kpc^3]$ & $R_2[\rm R_{500}]$ & $\beta$ \\
		\hline
        \gadgetx\ & $8.027 \times 10^4$ & 0.240 & $2.304 \times 10^5$ & 0.059 & 0.770\\
        \gadgetmusic\ & $2.773 \times 10^5$ & 0.045 & $8.372 \times 10^4$ & 0.210 & 0.751 \\
		\hline
		fitting function & \multicolumn{5}{c}{$\rho(R) = \sum_{i=1}^2 \rho_i \left[1 + \left(\frac{R}{R_i}\right)^2\right]^{-3\beta/2}$}\\ 
		\hline
	\end{tabular}
\end{table*}

\begin{figure}
    \centering
    \includegraphics[width=0.5\textwidth]{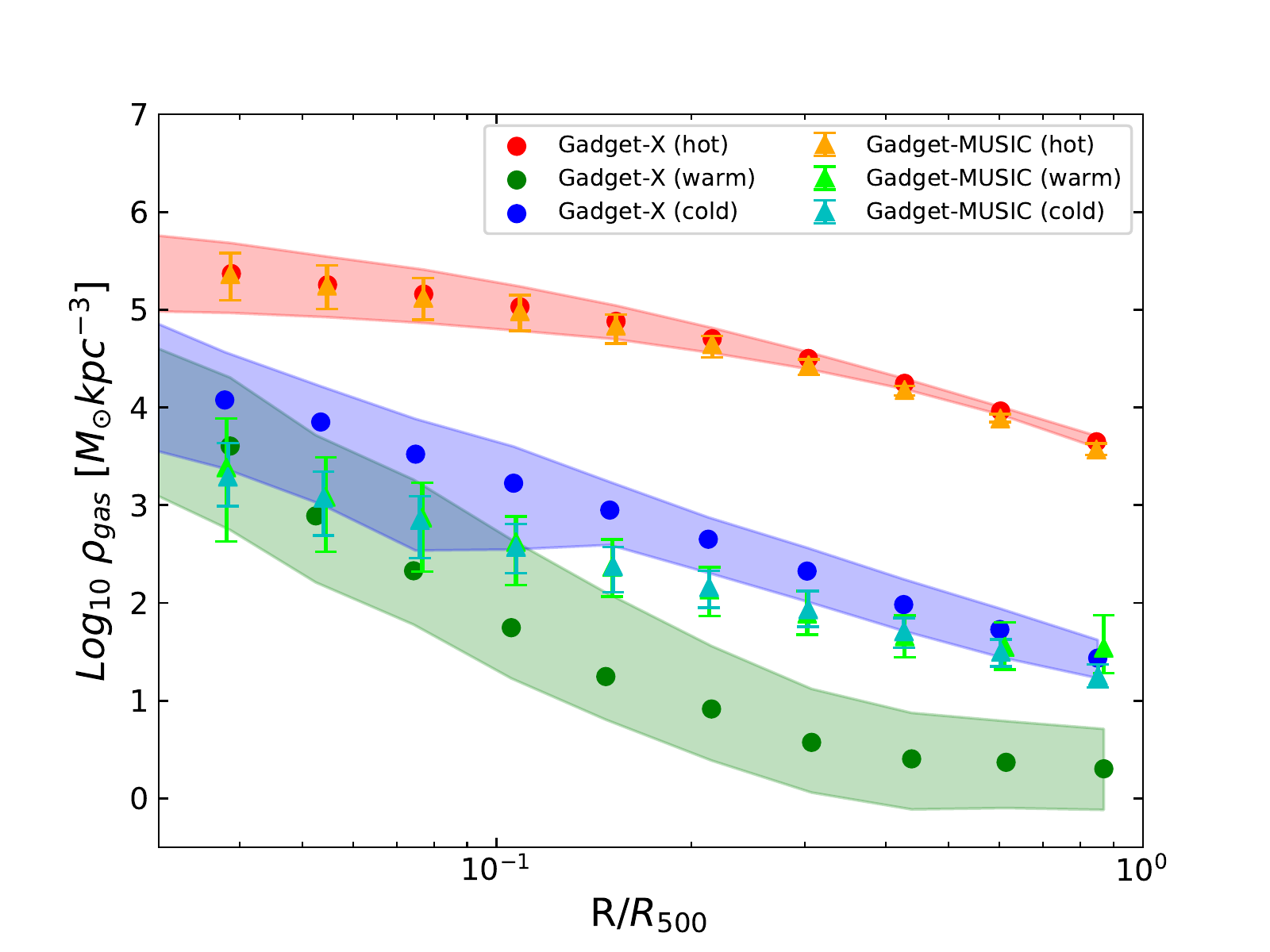} 
    \caption{Gas density profiles for gas at different temperatures (in the case of 3 temperature ranges, see text). Circles and triangles represent respectively \gadgetx\ and \gadgetmusic. Red (orange), green (lime) and blue (cyan) circles show hot, warm and cold gas median density profiles from \gadgetx\ (\gadgetmusic). Shaded regions (for \gadgetx) and  error bars (for \gadgetmusic) present the $16^{th}$ and $84^{th}$ percentiles of all cluster profiles.}
    \label{fig:gas_tempdensity}
\end{figure}

\begin{table}
	\centering
	\caption{The minimum, median and maximum spectroscopic-like temperatures of the clusters.}
	\label{tab:Tsl}
	\begin{tabular}{c c c c}
	    \hline
        Simulation & & $T_{500, sl}$ [keV] &  \\
		\hline
		 & min & median & max \\
		\hline
        \gadgetx\ & 2.68 & 5.67 & 12.56 \\
        \gadgetmusic\ & 2.46 & 4.92 & 9.94 \\
        \hline
	\end{tabular}
\end{table}

It is widely known that gas properties, such as density, temperature, pressure and entropy, show self-similar profiles \citep[see among other papers][]{Vikhlinin2006,Cavagnolo2009,Arnaud2010,Baldi2012,Planelles2017,Biffi2018,Ghirardini2019}. Moreover, it has been proposed that gas metallicity is homogeneously distributed near the outer radii of clusters \citep[see][for example]{Mantz2017,Biffi2017,Vogelsberger2018}. However, gas physical profiles (especially in cluster centres) seem to depend on the cluster dynamical state \citep[e.g.,][]{Lovisari2019} or on the CC/NCC classification \citep[e.g.,][]{Baldi2007,Leccardi2008}.

We present gas profiles from the two hydrodynamic simulations and also compare with observational results in this section. We note here that the gas profiles are calculated by summing over all the gas particles within the cluster. Because SAMs only provide gas properties within galaxies, which cannot be compared with observed results from X-ray telescopes, we do not include SAMs in this section. To compare with X-ray observational results of the smoothly distributed hot gas within the cluster, we select gas particles in the simulation with a temperature, T > 0.3 keV, and gas density, $\rho < 0.1 \,{\rm cm}^{-3}$ i.e., lower than the star forming threshold.
We calculate the gas temperature profile using the spectroscopic-like formula from \citet{Mazzotta2004}:
\begin{equation}
    T_{sl} = \frac{\Sigma_i m_i \rho_i T_i^{1/4}}{\Sigma_i m_i \rho_i T_i^{-3/4}},
\label{Tsl}
\end{equation}
where $m_i$ is gas mass, $\rho_i$ is gas density and $T_i$ is gas temperature of each considered gas particle. The distribution of the estimated gas temperatures in the clusters is shown in Table~\ref{tab:Tsl}.

For the metallicity, we consider the simplified emission-weighted formula:
\begin{equation}
    Z_{EW} = \frac{\Sigma_i m_i \rho_i \sqrt{T_i} Z_i}{\Sigma_i m_i \rho_i \sqrt{T_i}},
\label{Zew}
\end{equation}
The metallicity profile uses emission-weighted gas metallicity which is normalised to solar metallicity with respect to \citet{Asplund2009} ($Z_{\odot}=0.0134$).
To have a consistent comparison with our simulation data, only observational data on clusters  with $z$ < 0.1 and $M_{500} > 4.0 \times 10^{14}\,\Msun$ from observations are considered in this subsection. Finally, we normalize the profile with respect to $R_{500}$ as done in observations.

\subsubsection{Gas density profile}

Previous studies, using dark-matter-only simulations \citep{LeBrun2018}, full physics hydrodynamic simulations \citepalias{Mostoghiu2019} and observations at cluster scales \citep[e.g.][]{McDonald2017}, have revealed that both the total mass profile and the gas density profile show self-similar behaviours out to redshift $\sim 2$ in the outer region. We revisit this feature here by comparing with observed profiles to detail any differences and to understand the physics behind this behaviour. 

The gas density profile is presented in Fig. \ref{fig:gas_density0}, where we compare two hydrodynamic simulation results with the gas density profile from various observed data. Three clusters selected from \citet{Vikhlinin2006} with the criteria listed in the last paragraph of section \ref{subsec:gas} are shown by dashed lines: A478, A1795 and A2029; the median gas density profile at $0.0 < z < 0.1$ is taken from \citet{McDonald2017} which uses 27 clusters with masses spanning $4 \times 10^{14}\,\Msun < M_{500} < 1.2 \times 10^{15}\,\Msun$. This sample is X-ray flux-limited and constrained in redshift, as originally selected in \citet{Vikhlinin2009}. It has a similar number of CC, moderate CC and NCC clusters.

The top panel also shows the results of fitting the double $\beta$-model \citep{Mohr1999}. The residuals between the median and fits with four different models are presented in the bottom panel. The four models are respectively the double $\beta$-model \citep{Mohr1999}, the $\beta$-model \citep{Cavaliere1976}, the S$\rm{\Acute{e}}$rsic profile \citep{Sersic1963} and the NFW profile \citep{Navarro1997} shown with solid, dashed, dash-dotted and dotted lines for \gadgetx\ (red) and \gadgetmusic\ (blue). The S$\rm{\Acute{e}}$rsic and NFW fitting functions used here are mainly for a simple comparison to the distribution that is typically assumed for the stellar and the DM components.

The gas density profile is highly peaked towards the centre, which is in agreement with many observational results \citep[see][for example]{Pointecouteau2004}. And this is also the reason why the best fitting function is the double $\beta$-model. At outer cluster radii, the gas density profile between the hydro-simulations (\gadgetx\ and \gadgetmusic) and observations (three clusters A478, A1795 and A2029 from \citet{Vikhlinin2006} and \citet{McDonald2017}) shows a consistent trend, but larger discrepancies are present in the central region ($r \lessapprox 0.1 \times R_{500}$). Compared with the two hydro-simulations, the three individual clusters and \citet{McDonald2017} respectively give a higher and lower density. However, error bars also significantly increase in the cluster centre, indicating gas density changes between individual clusters. This inner scatter can be reduced by separating clusters into cool-core and non-cool-core or into dynamically relaxed and un-relaxed (see more details in the appendix \ref{app:1}). This means that the density profile is sensitive to both the implemented physical models and the detailed halo formation history. Overall, the agreements between two hydro-simulations and between the simulations and observational results are relatively good. This confirms the self-similarity of the gas profiles and further indicates that baryon models play a weak role in shaping the gas density profiles, especially in the outer regions, where gas follows the distribution of dark matter. Therefore, it is also not surprising to see that gas density profiles with much less scatter display a similar trend to the stellar density profile, even though they are much steeper in the outer regions.


\medskip

The total gas density profiles of the hot component of the clusters produced by \gadgetx\ and \gadgetmusic\ are very similar. We, then, investigate whether gas at different temperatures shows similar trends between the two simulations. We first separate the gas into hot (>$10^7$ K), warm ($10^7 - 10^5$ K) and cold (<$10^5$ K) phases \citep[see, for example,][for a similar definition]{Cui2019}. As shown in Fig.~\ref{fig:gas_tempdensity}, the gas mass is dominated by hot gas in clusters and our two simulations also share a similar profile for hot gas, even though \gadgetmusic\ is slightly lower at the outer radii compared to \gadgetx\ in agreement with Fig.~\ref{fig:gas_density0}. The warm and cold gas profiles from \gadgetmusic\ are very similar and also close to the cold gas profile from \gadgetx, all of which are about 2 orders of magnitude lower than the hot gas profile. The warm gas profile from \gadgetx\ is lower than the others, which may be because \gadgetx\ uses a much more efficient cooling rate (metal cooling) than \gadgetmusic\ (which has metal independent cooling). In the very inner cluster region all the different profiles converge towards $\rho_{\rm gas} \approx 10^{5.5}\,\Msun\,{\rm kpc}^{-3}$. 

\subsubsection{Gas temperature profile} \label{subsec:gas temp}

\begin{figure}
    \centering
    \includegraphics[width=0.5\textwidth]{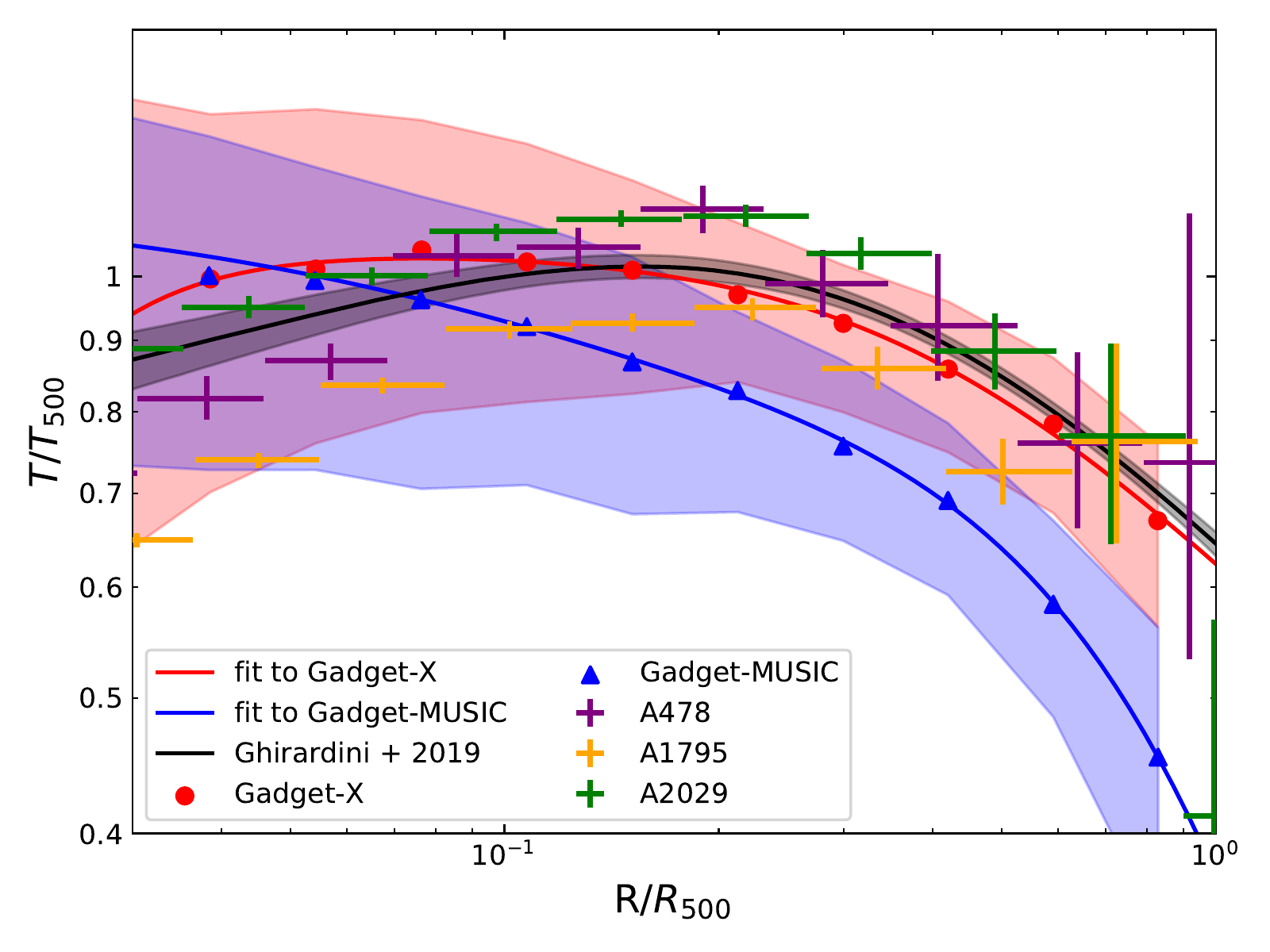}
    \caption{Radial spectroscopic-like gas temperature profile. Red circles and blue triangles respectively represent the median profile from \gadgetx\ and \gadgetmusic\ where the shaded areas show the $16^{th}$ and $84^{th}$ percentiles of all cluster profiles. We note here that an additional normalisation factor is applied to the two simulations; please refer to the text for details. Red and blue solid lines show fits of the formula from \citet{Ghirardini2019}. Purple, orange and green crosses respectively represent cluster A478, A1795 and A2029 temperature profiles obtained from \citet{Vikhlinin2006}. The black solid line with fitting percentile error shown as a shaded region is from \citet{Ghirardini2019}.}
    \label{fig:gas_temp}
\end{figure}

\begin{table*}
	\centering
	\caption{The fitting function from \citet{Ghirardini2019} and parameters for Fig. \ref{fig:gas_temp}.}
	\label{tab:temp_table}
	\begin{tabular}{c c c c c c c}
		\hline
		fitting parameters & $T_0\ [T_{500}]$ & $T_{\rm min}\ [T_{500}]$ & $r_{\rm cool}\ [R_{500}]$  & $a_{\rm cool}$ & $r_t\ [R_{500}]$ & c\\
		\hline
		\gadgetx\ & 1.046 & 0.212 & 0.016 & 3.170 & 0.393 & 0.515 \\
		\gadgetmusic\ & 0.759 & 1.120 & 0.099 & 1.242 & 1.141 & 2.574\\
		\hline
		fitting function & \multicolumn{6}{c}{$\frac{T(x)}{T_{500}} = T_0 \frac{\frac{T_{\rm min}}{T_0} + \left(\frac{x}{r_{\rm cool}}\right)^{a_{\rm cool}}}{1 + \left(\frac{x}{r_{\rm cool}}\right)^{a_{\rm cool}}} \frac{1}{\left(1 + \left( \frac{x}{r_t}\right)^2\right)^{\frac{c}{2}}}$} \\
		\hline
	\end{tabular}
\end{table*}

It is well known that the cluster gas temperature shows a tight scaling relation with its mass \citepalias[see][and references therein]{Cui2018}. Observations also suggested that the gas temperature profile shows similarity after rescaling with cluster mass \citep[e.g.][]{Vikhlinin2005,Rasmussen2007,Pratt2007,Baldi2012,Ghirardini2019}.
\citet{Ghirardini2019}, for example, rescaled the temperature profiles derived for the X-ray Cluster Outskirts Project \citep[X-COP,][]{Eckert2017} clusters as
\begin{equation}
    T_{500, G+19} = 8.85 keV \left(\frac{M_{500}}{10^{15}\ h^{-1}_{70}M_\odot}\right)^{2/3} E(z)^{2/3} \frac{\mu}{0.6},
    \label{T500G}
\end{equation}  
where $E(z)$ is defined as $E^2(z) = \Omega_m (1+z)^3 + \Omega_{\Lambda}$ and $\mu$ is the mean molecular weight per gas particle, which \citet{Ghirardini2019} assumed to be equal to 0.6125 \citep{Anders1989}. Consistent with \citet{Ghirardini2019}, we adopt the same $\mu$ for calculating the gas temperature $T_{sl}$ from the simulations and use $T_{500, G+19}$ for the normalisation.
However, we find that $T_{500, G+19}$ from equation~\ref{T500G} with the $M_{500}$ from our simulation is a little higher than $T_{500, sl}$. That could be due to the fact that $M_{500}$ in $T_{500, G+19}$ is based on the hydrostatic equilibrium assumption, which tends to give a lower value. As shown in \citetalias{Cui2018}, there is a slight (about 14 per cent) off-set in the $M_{500} - T_{500}$ relation between the hydro-simulations and observation. 
We make a detailed investigation of the difference between $T_{500, sl}$ and $T_{500, G+19}$ and find a similar deviation, which is presented in appendix B1. Therefore, we apply this correction fraction to our simulation result in Fig.~\ref{fig:gas_temp}.

In Fig.~\ref{fig:gas_temp} the gas temperature profiles are compared, with A478, A1975 and A2029 from \citet{Vikhlinin2006} shown by data points with error bars and the fitting result from \citet{Ghirardini2019} shown by a solid black line plus a grey shaded region. These three clusters from \citet{Vikhlinin2006} adopt the same normalisation with $T_{500, G+19}$. The profile by \citet{Ghirardini2019} is based on a total number of 12 clusters which are originally selected from the first Planck Sunyaev-Zel'dovich (SZ) catalogue \citep{Planck2014} with a SZ signal limitation and low redshift (0.04 < z < 0.1). This sample has mass range $3 \times 10^{14}\,\Msun < M_{500} < 9 \times 10^{14}\,\Msun$ with 4 CC and 8 NCC based on the central entropy value measured by \citet{Cavagnolo2009}.


Observed temperature profiles show a slightly increasing trend from the cluster centre outwards to r $\sim 0.2 \times R_{500}$ and decrease from 20 percent of $R_{500}$ to the outer regions. The gas temperature profile from \gadgetx\ is in good agreement with observations, while \gadgetmusic\ has a much higher and flatter profile in the cluster centre compared to observations. Its profile in the outer regions is also lower and slightly steeper than the observed results. This could be caused by the normalisation. Unlike the gas density profile, the gas temperature profile is strongly affected by the baryon models. Overall, there are large scatters in these two hydro-simulations, especially in the cluster centre region. We further test whether this scatter is caused by cluster dynamical state and CC/NCC classification in appendix \ref{app:1} and find that the scatter in the cluster centre regions can be significantly reduced by separating the clusters into CC/NCC.


\subsubsection{Gas metallicity profile}

\begin{figure}
    \centering
    \includegraphics[width=0.5\textwidth]{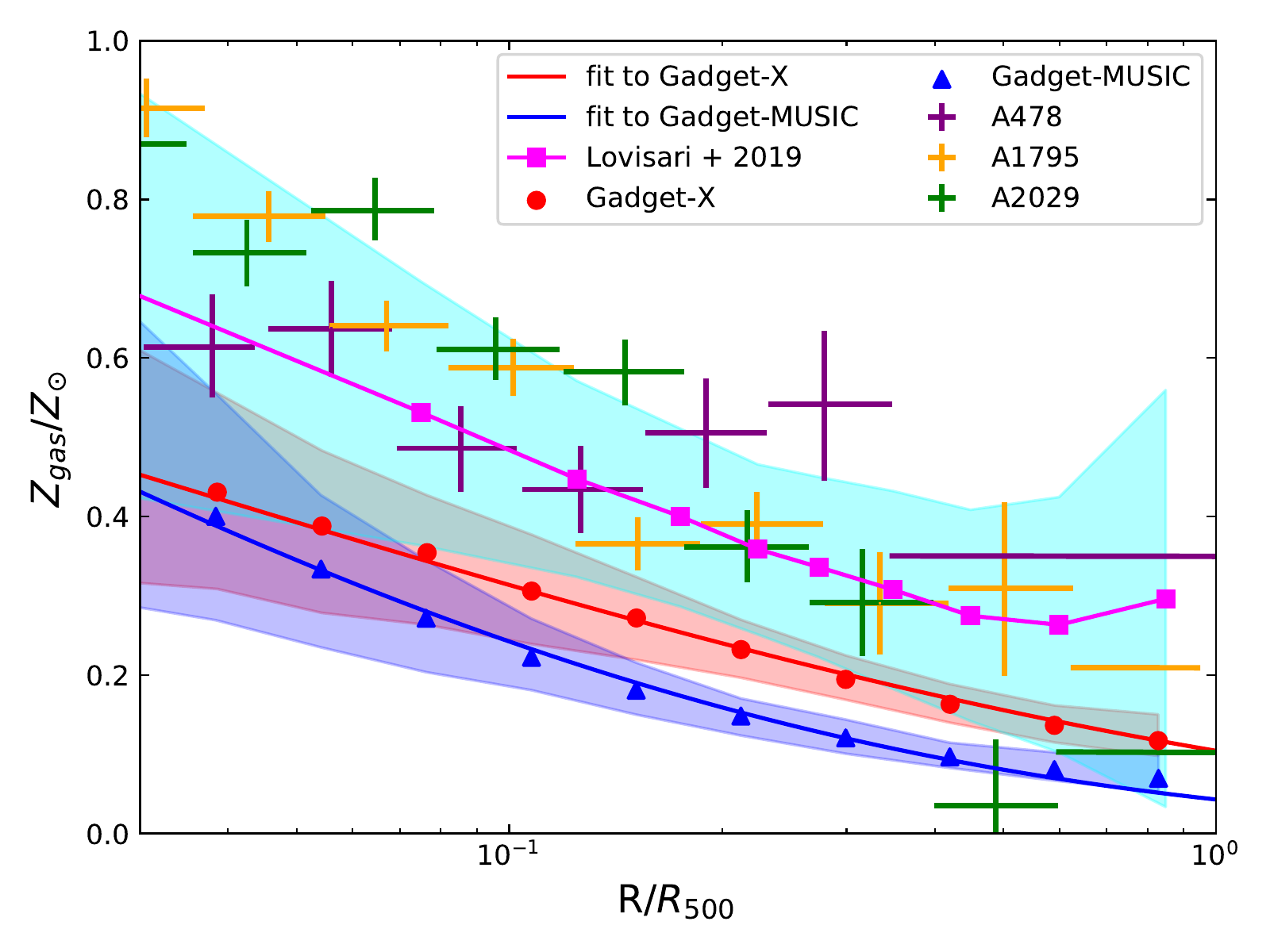}
    \caption{Radial emission-weighted gas metallicity profile. The red circles and blue triangles respectively represent \gadgetx\ and \gadgetmusic\ where solid lines and  shaded areas show the fits with a S$\rm{\Acute{e}}$rsic profile and the $16^{th}$ and $84^{th}$ percentiles of all cluster profiles, respectively. The purple, orange and green crosses show the gas metallicity profile of the galaxy clusters: A478, A1795 and A2029 respectively, taken from \citet{Vikhlinin2005}. The magenta rectangles and cyan region show the median abundance profile and scatter of measurements given by \citet{Lovisari2019}.}
    \label{fig:gas_metal}
\end{figure}

\begin{table}
	\centering
	\caption{The fitting function and parameters for Fig. \ref{fig:gas_metal}.}
	\label{tab:gasz_table}
	\begin{tabular}{c c c}
		\hline
		fitting parameters & \gadgetx & \gadgetmusic\\
		\hline
		$Z_0[Z_{\odot}]$ & 0.602 & 0.669 \\
		$R_0[R_{500}]$ & 0.001 & 0.002\\
		b & 0.419 & 0.863 \\
		\hline
		fitting function & \multicolumn{2}{c}{$Z(R)\ =\ Z_0e^{-b[\left(\frac{R}{R_0}\right)^{\frac{1}{4}} - 1]} $} \\
		\hline
	\end{tabular}
\end{table}

The metal enrichment of the ICM involves numerous astrophysical processes, such as stellar nucleosynthesis and supernova explosions. The resulting metals will enrich the surrounding ICM thanks to multi-scale mixing processes, such as galactic winds, AGN feedback, ram-pressure stripping and mergers. Both observation \citep[e.g.][]{Leccardi2008, Werner2013, McDonald2016, Mantz2017} and simulations \citep[e.g.][]{Biffi2017,Biffi2018,Vogelsberger2018} have suggested a homogeneous distribution of metals in space and time. The uniform metal distributions in the outskirts of nearby clusters indicates an early enrichment of the ICM, most of which takes place before cluster formation. 

Fig.~\ref{fig:gas_metal} shows the emission-weighted gas metallicity profile as a function of radius normalised to $R_{500}$. The metallicity profile is compared with the stacked profile by \citet{Lovisari2019} who study galaxy cluster metallicity profiles using a sample of 207 nearby galaxy groups and clusters observed with XMM-Newton. The stacked profile is estimated with a Monte Carlo method based on performing 10,000 realisations of the profiles by randomly varying the observational data points of the metallicity profile. Besides this stacked dataset we also compare to three individual cluster results from \citet{Vikhlinin2006}. 

In general, the radial gas metallicity profile is centrally peaked and gradually decreases from the centre to 0.2-0.3 $R_{500}$, where it flattens and stays almost constant out to large radii. Both simulations and observations follow the same trend. However, the simulations are slightly lower (much lower metallicity in \gadgetmusic\ than \gadgetx) than the observed data points at the outer radii. As shown in \cite{Rasia2008}, the metallicity derived from XMM-Newton spectra of simulated mock observations was generally in good agreement with the emission-weighted metallicity. So this difference must arise for other reasons. For \gadgetx\, we especially compare to \cite{Biffi2018} which showed a good match to the observational results and find that this lower metallicity is mainly caused by a lower star formation and inefficient kinetic SN feedback in the \gadgetx\ run, therefore less metal is produced. The lower star formation is introduced by several parameter changes compared to \cite{Biffi2018}, driven mostly by the choice of a much larger gravitational softening length. Furthermore, as indicated in \citet{Vogelsberger2018}, this could be a resolution issue, which is also consistent with our previous findings. Because of the modest resolution of these simulations, the star formation starts later in time leading to a reduced amount of metals in both stars and diffuse gas. Similarly to the observed profile \citep[see also][]{Elkholy2015,Mernier2016}, gas metallicity becomes basically flat at $r > R/R_{500}$. The error bars of simulated profiles for both \gadgetx\ and \gadgetmusic\ are very narrow at outer radii: this indicates that the variation of gas metallicity between clusters is very small and that the detailed cluster formation history has a limited effect on the metal enrichment of the recent accreted ICM. The AGN feedback from \gadgetx\ seems to have a non-negligible effect by boosting gas metallicity at larger radii compared to \gadgetmusic\ as found in \citet[]{Rasia2015, Biffi2018}. We further fit the metallicity profile from the two simulations with the S$\rm{\Acute{e}}$rsic profile, which provides a very good match to the simulation data. The fitting results are listed in Table~\ref{tab:gasz_table}.

\subsection{SFR and sSFR distributions} \label{secsub:sfr}

\begin{figure*}
    \centering
    \includegraphics[width=1.\textwidth]{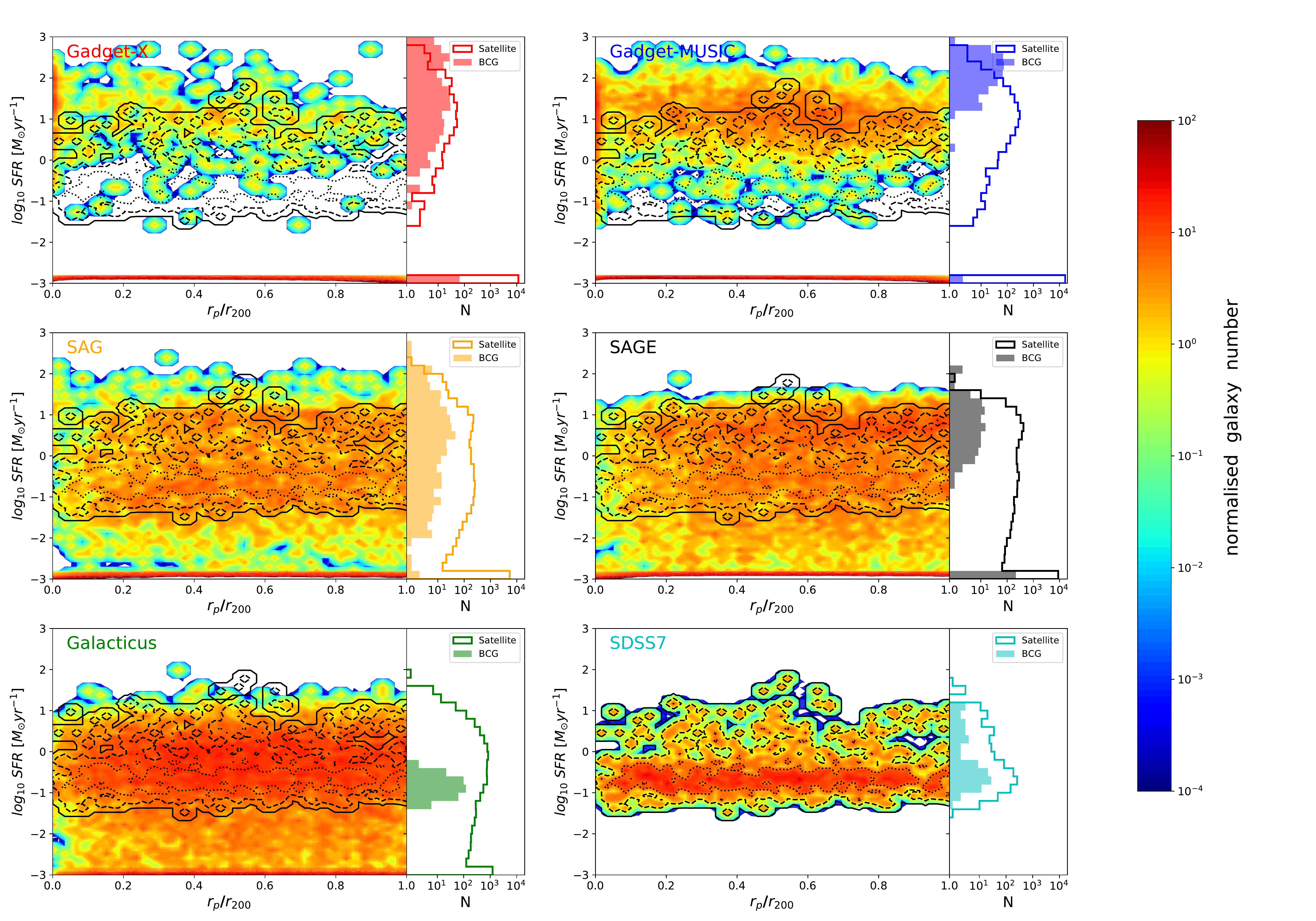}
    \caption{The distribution of satellite-galaxy SFR with radius inside clusters. No BCGs are included in the left-hand panels. The colour bar shows the normalised galaxy number density with respected to total cluster number. The SFR histograms for BCGs (filled) and satellite galaxies (solid) are shown in the right-hand sub-panels. Solid, dashed and dotted black lines respectively represent the $16^{th}$, $50^{th}$ and $84^{th}$ percentiles of the pixel value from SDSS result, which is repeated in all panels for comparison. A Gaussian kernel is used to smooth the pixels. All galaxies with SFR less than 0.001\,$\Msun yr^{-1}$ are put in the lowest SFR bin.}
    \label{fig:SFR}
\end{figure*}

\begin{figure*}
    \centering
    \includegraphics[width=1.\textwidth]{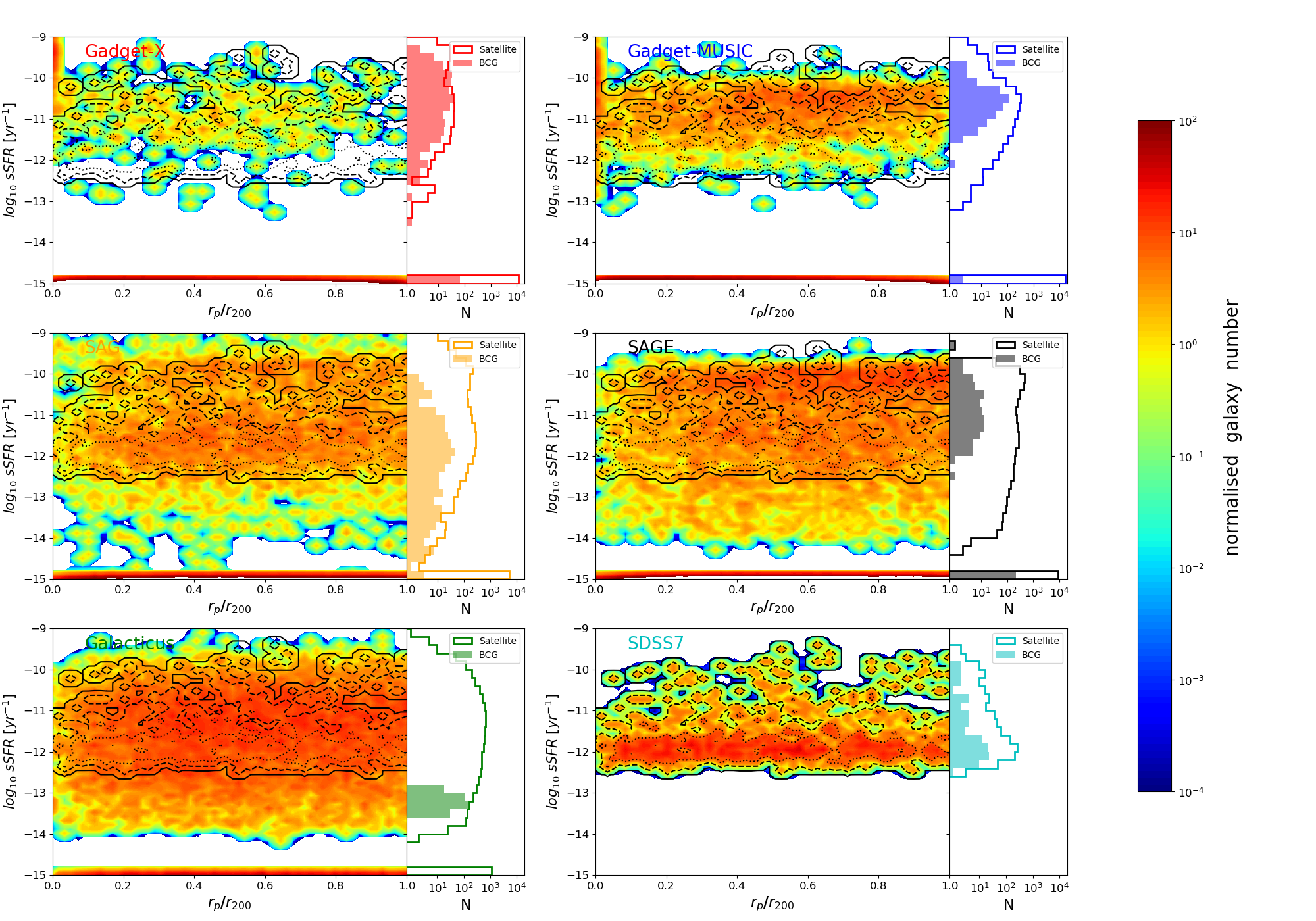}
    \caption{Similar to the SFR distribution in Fig.~\ref{fig:SFR} but for the sSFR distribution of satellite galaxies inside the clusters. The BCG distributions are only included in the right-hand panels as filled steps. All galaxies with sSFR less than $10^{-15} yr^{-1}$ are put in the lowest sSFR bin.}
    \label{fig:sSFR}
\end{figure*}

The star formation rate (SFR) serves as a connection between gas and stars. It also determines galaxy colour and connects to many galaxy properties such as shape and age. Many studies have revealed that the number fraction of red, quiescent (blue star-forming) galaxies decreases (increases) with increasing halo-centric radius in the cluster \citep[see][for example]{Weinmann2006,Bai2009}. Using 69 local clusters, \citet{Chung2011} claimed that the sSFR increases with projected radius after applying a SFR cut ($L_{IR} > 4.7 \times 10^{10} L_{\odot}$), which is consistent with the study of $H_{\alpha}$ star-forming galaxies by \citet{Lewis2002}. However, \citet{Lagana2018} studied 17 galaxy clusters at intermediate-to-high redshifts ($0.4 < z < 0.9$) and found that the mean SFR of both SF galaxies and quenched galaxies does not change with cluster-centric radius. Similar results for clusters at higher redshifts ($1.0 < z < 1.75$) were also found by \citet{Alberts2016}. It is interesting to see whether the modelled galaxies are in agreement with these findings at lower redshifts.

The galaxy SFR is directly provided by SAMs, while the galaxy SFR from hydro-simulations is calculated by summing up the instantaneous SFR from gas that lies within twice its stellar half-mass radius. We also compute the galaxy SFR via its mass change from the previous snapshot divided by the time difference between the two snapshots which is about 300 Myr. Both methods give very similar SFR values. Therefore, we use the galaxy SFR from the first method here. The SFRs of SDSS galaxies are estimated through their H$\alpha$ luminosities \citep{Brinchmann2004}. Rather than firstly separating the star-forming galaxies from quenched galaxies with some arbitrary values, we divide the data into 60 bins in both radius and SFR or sSFR. Any galaxy with SFR value below $0.001\,\Msun\,{\rm yr}^{-1}$ is set at $0.001\,\Msun\,{\rm yr}^{-1}$; meanwhile, its sSFR is set at $10^{-15}{\rm yr}^{-1}$. Each pixel value is normalised by the total cluster number. Finally, a Gaussian filter with standard deviation 0.5 pixels is used to smooth both the SFR and the sSFR 2D images.  

In Fig.~\ref{fig:SFR} and Fig.~\ref{fig:sSFR}, we show respectively the distribution of galaxy SFR and sSFR for satellite galaxies. Each panel shows one model with the colour-bar indicating a normalised galaxy number density. Solid, dashed and dotted black lines respectively represent the $16^{th}$, $50^{th}$ and $84^{th}$ percentiles of pixel value from the SDSS results. These lines are repeated in each panel for reference. We also show histograms for the BCG (only presented in the histogram with filled steps) and satellite galaxy SFRs in the right-hand sub-panels. The SFR of the BCGs from the hydrodynamic simulations is defined as the total SFR from all gas particles within $0.015 \times R_{\rm 200}$, while the sSFR of the BCGs is calculated by ${\rm SFR}/M_{\rm star}$ within $0.015 \times R_{\rm 200}$. We find only a few clusters from \gadgetmusic\ do not contain gas particles within $0.015 \times R_{\rm 200}$. We again test that extending this limiting radius to $0.05 \times R_{\rm 200}$ and find that only \gadgetmusic\ has its BCG's SFR dropping slightly. 

For galaxy SFR, it is clear that the most galaxies have a SFR less than 1\,$\Msun\,{\rm yr}^{-1}$ for all models and SDSS result. However, the detailed SFR distributions differ between these models and SDSS galaxies: there is no observed galaxy with SFR < $0.01\,\Msun\,{\rm yr}^{-1}$, although this could reflect observational limitations. All of the modelled galaxies have a significant fraction with SFR < $0.001\,\Msun\,{\rm yr}^{-1}$; for the modelled galaxies with SFR > $0.001\,\Msun\,{\rm yr}^{-1}$, SAMs have much smoothed distribution filling the whole plot and tend to have a better agreement to the SDSS distribution, while the two hydro-simulations present more star-forming galaxies and fewer galaxies between $1\,\Msun\,{\rm yr}^{-1} > SFR > 0.001\,\Msun\,{\rm yr}^{-1}$. Overall, the radial distribution of the galaxy SFR is more or less constant, which seems to be in agreement with the high redshift results \citep{Lagana2018,Alberts2016} rather than the low redshift results \citep{Weinmann2006,Bai2009,Chung2011}. If the star-forming galaxies are selected out from SDSS with SFR > $1\,\Msun\,{\rm yr}^{-1}$, we can see their number fractions decrease towards the cluster centre. In agreement with the SFR distribution, the sSFR distribution also does not show much evolution with radius. Again, the contours from the two hydrodynamic simulations seem a little higher than SDSS result, while \sage\ and \galacticus\ have fewer galaxies with and a higher sSFR compared to the SDSS result. This indicates that high SFR galaxies in \sage\ and \galacticus\ are too massive as compared to SDSS galaxies. From the histogram of BCG SFR, SDSS seems to have a double peak with the majority of BCGs having a quenched SFR $< 1\,\Msun\,{\rm yr}^{-1}$. Apart from this the BCGs from \galacticus\ are in good agreement with SDSS, all the other models have their BCG SFR widely spread and the BCGs in the two hydrosimulations tend to have a higher SFR than their star forming galaxies. 


\section{Conclusions}

We have studied the physical profiles of galaxy clusters using 324 massive clusters from two hydrodynamic simulations and modelled by three semi-analytic models at $z = 0$. Observations including massive galaxy groups from the SDSS catalogue and many X-ray datasets are used to compare galaxy and gas properties of our models with reality. For stellar properties, we consider galaxy stellar density, galaxy age and metallicity after applying the same mass cut for both modelled and SDSS galaxies. For gas properties, we study the gas density, temperature and metallicity. We also further investigate SFR and sSFR distributions as a function of halo-centric radius. The followings are our main conclusions.

Regarding self-similarity, the gas density profiles from both sets of simulated clusters present the highest level of similarity and are characterised by very small scatter at outer radii. There is a slightly larger scatter in the cluster centre indicating that different physical states of the cluster play a role, such as dynamical state or CC/NCC dichotomy. 

The simulated gas metallicity profiles also have a very small (similar) scatter at the cluster outer radii. This scatter is smaller than what is typically observed and could be due to systematic effects in the metallicity measurements. However, the normalisation of the simulated metallicity profiles tends to be lower than the observed results. 

Both gas temperature profiles and cumulative stellar mass profiles display a relatively large scatter, implying that these quantities are more influenced by the detailed halo formation history. By separating the clusters into relaxed and un-relaxed groups, we further show in appendix \ref{app:1} that relaxed clusters tend to have a higher (lower) temperature (stellar density) profile compared to un-relaxed ones. Furthermore, not surprisingly, the CC clusters have a lower temperature profile in the cluster centre than NCC clusters.

Galaxy age, metallicity and (s)SFR do not present any clear radial distributions and have a large scatter at a given radius. Therefore, they are presented in a 2D map for detailed investigation and comparison. They do not show any clear sign of radial dependence.

In general, it is important to compare the simulated cluster results with observations in order to understand whether the baryon models in the simulations impact on the shape and normalization of the profiles. The gas density profiles present the best agreement between different models and observations. This means that the gas density at outer radii ($R > 0.2 \times R_{500}$) depends only weakly on the baryon models.
\gadgetx\ simulated clusters have a median temperature profile that agrees with observation with a correct shape, while \gadgetmusic\ show a higher and flatter temperature profile in the cluster centre with a steeper and lower temperature profile at large radii compared to observation. 
The modelled gas metallicity profiles are slightly lower than the observed ones with \gadgetx\ closer to the observed results at middle and outer radii and \gadgetmusic\ closer in the innermost region. Several reasons might generate this shift including the modest resolution of our sample which could delay star formation and therefore metal production.

For the cumulative stellar mass profile, the three SAMs similarly present a good match to the SDSS result. However, both \gadgetx\ and \gadgetmusic\ show much higher (about 2 and 8 times respectively) stellar profiles than the one from SDSS. It seems that we need even stronger feedback to match the SDSS result.

We also find some discrepancies between models and observation when comparing the 2D distributions of galaxy age, metallicity and (s)SFR: The BCGs from the observed clusters and from the SAMs have a relatively old age, which is not seen in both hydro-simulations; observed galaxies tend to have higher metallicity than the models; both SFR and sSFR show similar distributions between SAM modelled galaxies and the observed clusters for the galaxies with SFR > 0.1\,$\Msun\,{\rm yr}^{-1}$. However, the two hydro-simulations show larger deviations from the observations. Besides \galacticus, the BCG SFR from all the other models tend to have a significant fraction with values greater than 1. Finally, it is hard to draw a firm conclusion because, on one hand, the different theoretical models do not present consistent results and on the other hand, there is a large uncertainty in the quantities derived from observations.


There are two remaining questions that are not answered in the previous investigations: (1) what causes the scatter in these physical profiles? (2) what is the redshift evolution of these profiles during the formation of clusters? For the first one, we only partly study the effect of the dynamical state (and CC/NCC separation) of these clusters on the stellar density, gas density, temperature and metallicity profiles in appendix~\ref{app:1}. It is known that the dynamical state correlates with the cluster formation time \citep[see][for example]{Mostoghiu2019} which indeed plays a role in the scatter of galaxy profiles (see appendix~\ref{app:1} for details). However, to fully understand the physics behind this, we need census and correlation studies with halo properties, which will be detailed in a following work. For the second question, we will trace the halo progenitors and investigate their physical profiles at different redshifts. This is also planned for a further forthcoming paper.

\section*{Acknowledgements}

The authors thank the referee for his/her valuable suggestions and helps.
The authors sincerely thank Giuseppe Murante, Stefeno Borgani and Klaus Dolag for their works on the \gadgetx\ development, Johan Comparat and Claudia Maraston for their kind helps and useful discussions on the SDSS firefly catalogue, Veronica Biffi for discussions on gas profiles, Vittorio Ghirardini for his kind helps on the temperature profiles and Lorenzo Lovisari for his helps on the gas metallicity profile. The authors also thank Yang Lei for the calculation of K and E correction of SDSS galaxies and Robert Mostoghiu for useful discussions. 

This work has made extensive use of the Python packages --- Ipython with its Jupyter notebook \citep{ipython}, NumPy \citep{NumPy} and SciPy \citep{Scipya,Scipyb}. All the figures in this paper are plotted using the python matplotlib package \citep{Matplotlib}. This research has made use of NASA's Astrophysics Data System and the arXiv preprint server. 

The CosmoSim database used in this paper is a service by the Leibniz-Institute for Astrophysics Potsdam (AIP). The MultiDark database was developed in cooperation with the Spanish MultiDark Consolider Project CSD2009-00064.

The authors gratefully acknowledge the Gauss Centre for Supercomputing e.V. (www.gauss-centre.eu) and the Partnership for Advanced Supercomputing in Europe (PRACE, www.prace-ri.eu) for funding the MultiDark simulation project by providing computing time on the GCS Supercomputer SuperMUC at Leibniz Supercomputing Centre (LRZ, www.lrz.de).

WC \& JAP acknowledge supported from the European Research Council under grant number 670193. AK and GY are supported by the {\it Ministerio de Econom\'ia y Competitividad} and the {\it Fondo Europeo de Desarrollo Regional} (MINECO/FEDER, UE) in Spain through grant AYA2015-63810-P and MICIU/FEDER through grant PGC2018-094975-C21. AK further acknowledges funding through the Spanish Red Consolider MultiDark FPA2017-90566-REDC and thanks Blumfeld for l'etat et moi.
XY is supported by the National Science Foundation of China (NSFC, grant Nos. 11890692, 11833005, 11621303) and the 111 project No. B20019.

As requested by `The Three Hundred' policy, we state the contribution by each author to this paper: Q.L., W.C. and X.Y. formed the core team to do the analysis, produce the plots and write the paper; W.C. developed and supervised the project; E.R. provided critical feedback, contributed to the CC/NCC data and the interpretation of the results; R.D., M.D.P., J.P. and F.P. provided critical feedback for this work. W.C. and G.Y. contributed to the simulated data set. A.K. contributed to the halo catalogue. The author list is in alphabetic order with the exclusion of the first four authors.




\bibliographystyle{mnras}
\bibliography{paper}



\appendix

\section{The effects of cluster dynamical states and CC/NCC dichotomy}

The profiles investigated in this paper such as the gas density profile show a very strong self-similarity, i.e. very small error bars in the profile. While some have relatively large  error bars, such as the stellar density and gas temperature profiles. \citetalias{Mostoghiu2019} showed that the total density profile depends on the cluster dynamical state, i.e. the relaxed clusters have a higher scale radius $r_s$ than un-relaxed clusters, which essentially links with the halo formation time. Therefore, we naively expect that the cluster dynamical state is also responsible for the large dispersion in the physical profiles. In this section, we separate the clusters into relaxed and un-relaxed \citep{Cui2017}, CC and NCC \citep{Rasia2015} and try to understand whether cluster dynamical state and CC/NCC dichotomy are the main cause of the scatter in these profiles. The criteria for separating relaxed and un-relaxed clusters are based on three indicators: the virial ratio $\eta$, the centre-of-mass offset $\Delta_r$ and the fraction of mass in subhaloes $f_s$. We adopt the same limitations as \cite{Cui2017} to select dynamically relaxed clusters: $0.85 < \eta < 1.15$, $\Delta_r < 0.04$ and $f_s < 0.1$. While the CC and NCC clusters are separated by measuring the shape and level of the entropy profiles in the cluster central regions: the pseudo entropy $\sigma$ and the central entropy $K_0$. The CC clusters are selected with $\sigma < 0.55$ \citep{Rasia2015} and $K_0 < 60\ keV cm^2$ \citep{Cavagnolo2009}. 

\subsection{The effects of the cluster dynamical states on stellar density profile} \label{app:1}

\begin{figure*}
    \centering
    \includegraphics[width=\textwidth]{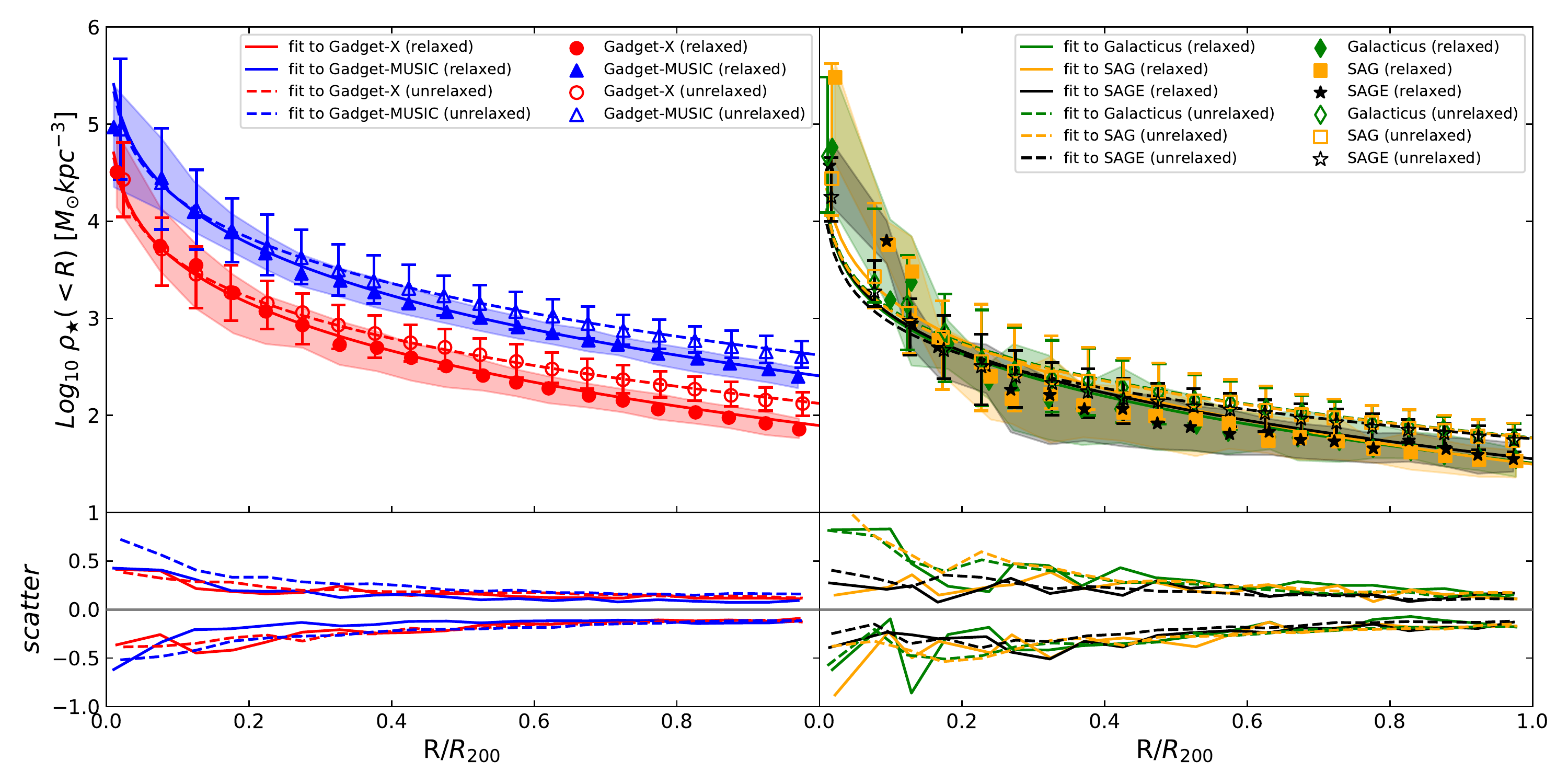}
    \caption{Cumulative stellar density profile for relaxed clusters and un-relaxed clusters. Left panel shows the results from the two hydro-simulations, while right panel is for the three SAMs. Shaded areas and  error bars show the $16^{th}$ and $84^{th}$ percentiles for relaxed and un-relaxed clusters respectively, while lower panels highlight their scatters. Solid and dashed lines respectively indicate the fitting for relaxed and un-relaxed clusters, using the formula from \citet{Lokas2001}. The fitting results are shown in table~\ref{tab:stellar_density4_state}.}
    \label{fig:stellar_density4_state}
\end{figure*}

\begin{table*}
	\centering
	\caption{The NFW fitting function and parameters for the cumulative density profiles in Fig.~\ref{fig:stellar_density4_state}. The fitting process is the same as for Fig. \ref{fig:stellar_density4}. We note here that only c is the free parameter in the fitting function.}
	\label{tab:stellar_density4_state}
	\begin{tabular}{c c c c c c c c}
        \hline
        fitting parameters & c & $10^{-3}R_{200}[\rm kpc]$ & $10^{-12}M_{200}[\Msun]$& & c & $10^{-3}R_{200}[\rm kpc]$ & $10^{-12}M_{200}[\Msun]$\\
        \hline
        cluster dynamical state & \multicolumn{3}{c}{relaxed} & & \multicolumn{3}{c}{un-relaxed} \\
		\hline
		Model \\ \hdashline
        \gadgetx\ & 2.880 & 2.204 & 3.524 & & 1.478 & 2.261 & 6.437\\
        \gadgetmusic\ & 4.285 & 2.221 & 11.723 & & 4.991 & 2.260 & 20.209\\
		\galacticus\  & 1.687 & 2.196 & 1.431 & & 2.589 & 2.253 & 2.820\\
		\sag\  & 3.017 & 2.196 & 1.394 & & 0.832 & 2.253 & 2.795\\
		\sage\ & 1.571 & 2.196 & 1.585 & & 0.453 & 2.253 & 2.747\\
		\hline
        fitting function & \multicolumn{7}{c}{$\rho(s) = M_{200} g(c)\left(\ln{(1+cs)}-\frac{cs}{1+cs}\right)/\left(\frac{4}{3}\pi s^3r_{200}^3\right)$} \\
        \hline
	\end{tabular}
\end{table*}

In Fig.~\ref{fig:stellar_density4_state}, we show the stellar density profiles from relaxed (filled symbols and solid lines) and un-relaxed clusters (open symbols and dashed lines). We compare the two hydro-simulations in the left panel and the three SAMs in the right panel for a better visualization. The relaxed clusters tend to have a shallower profile than the un-relaxed clusters, especially at outer radii. There is almost no difference in the cluster centre region for the two hydro-simulations, but the stellar density is slightly higher for the relaxed clusters in the SAMs. However, the scatter is basically at the same level for the relaxed and un-relaxed clusters: $\sim 0.5 - 0.1$ dex for the hydro-simulations and for the SAMs. The difference between the relaxed and un-relaxed cluster profiles indicates that the dynamical state has an impact on the stellar density profiles.

\subsection{The effects on gas density profile}
\begin{figure}
    \centering
    \includegraphics[scale = 0.5]{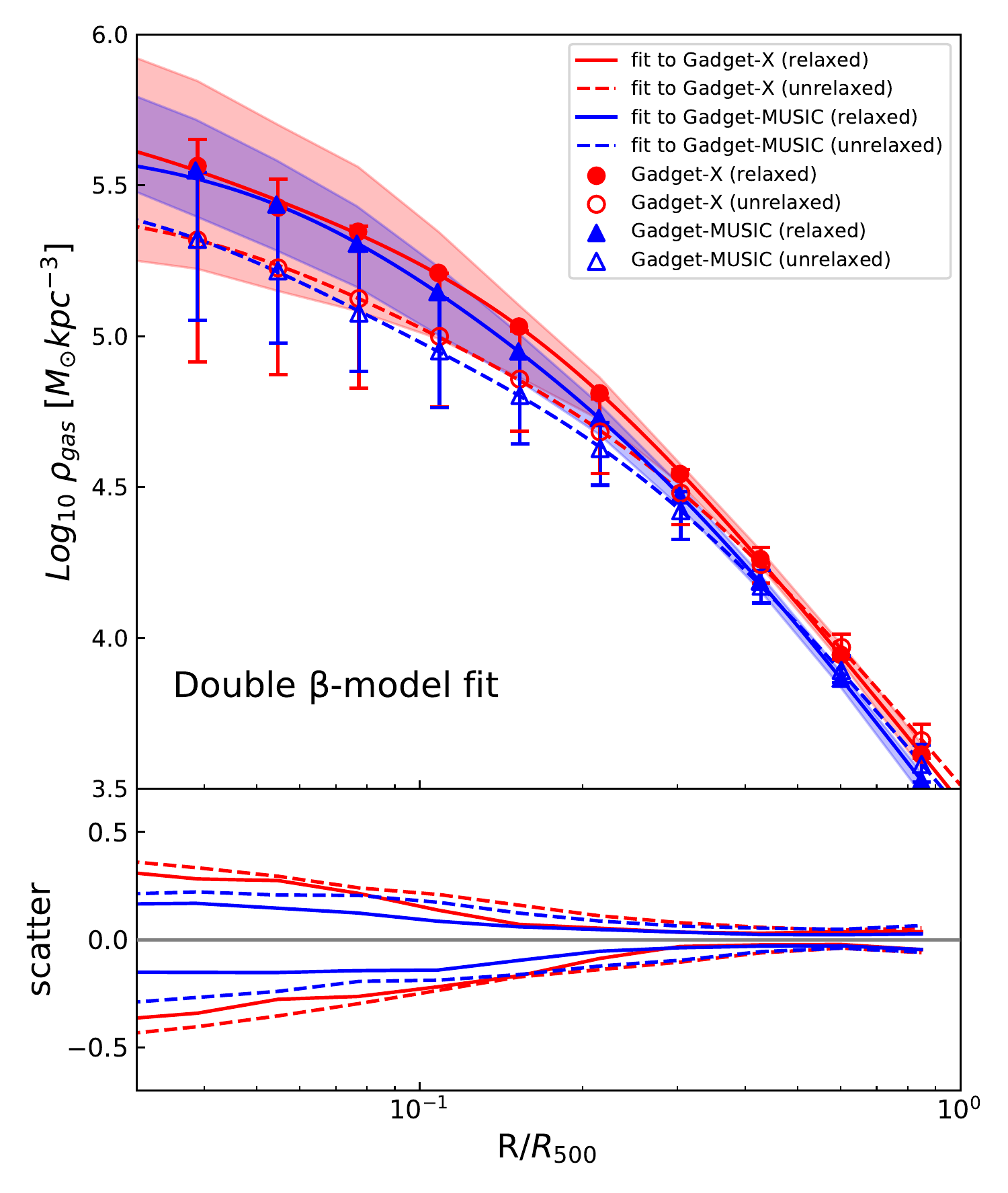}
    \caption{Radial gas density profile for relaxed clusters and un-relaxed clusters from the two hydro-simulations. Upper panel presents the results with shaded areas and  error bars indicating the $16^{th}$ and $84^{th}$ percentiles for relaxed and un-relaxed clusters respectively, while the lower panel highlights these scatters. Solid and dashed lines respectively indicate the fitting for relaxed and up-relaxed cluster with double $\beta$-model.}
    \label{fig:gas_density_state}
\end{figure}

\begin{table*}
	\centering
	\caption{The double $\beta$-model fitting function (based on \citet{Mohr1999}) and the fitting parameters for Fig. \ref{fig:gas_density_state}.}
	\label{tab:gas_density_state}
	\begin{tabular}{c c c c c c c}
	    \hline
        \multicolumn{2}{c}{fitting parameters}  & $\rho_1[\Msun/kpc^3]$ & $R_1[\rm R_{500}]$ & $\rho_2[\Msun/kpc^3]$ & $R_2[\rm R_{500}]$ & $\beta$ \\
		\hline
		model & dynamical state \\
		\hdashline
		\multirow{2}{*}{ \gadgetx\ } & relaxed & $3.679 \times 10^5$ & 0.034 & $2.291 \times 10^5$ & 0.143 & 0.757  \\
		& un-relaxed & $2.038 \times 10^5$ & 0.059 & $7.434 \times 10^4$ & 0.239 & 0.742 \\
		\hdashline
		\multirow{2}{*}{ \gadgetmusic\ } & relaxed & $3.606 \times 10^5$ & 0.071 & $7.259 \times 10^4$ & 0.220 & 0.812  \\
	    & un-relaxed & $2.539 \times 10^5$ & 0.046 & $7.348 \times 10^4$ & 0.225 & 0.748  \\
		\hline
		fitting function & \multicolumn{6}{c}{$\rho(R) = \sum_{i=1}^2 \rho_i \left[1 + \left(\frac{R}{R_i}\right)^2\right]^{-3\beta/2}$}\\ 
		\hline
	\end{tabular}
\end{table*}

\begin{figure}
    \centering
    \includegraphics[scale = 0.5]{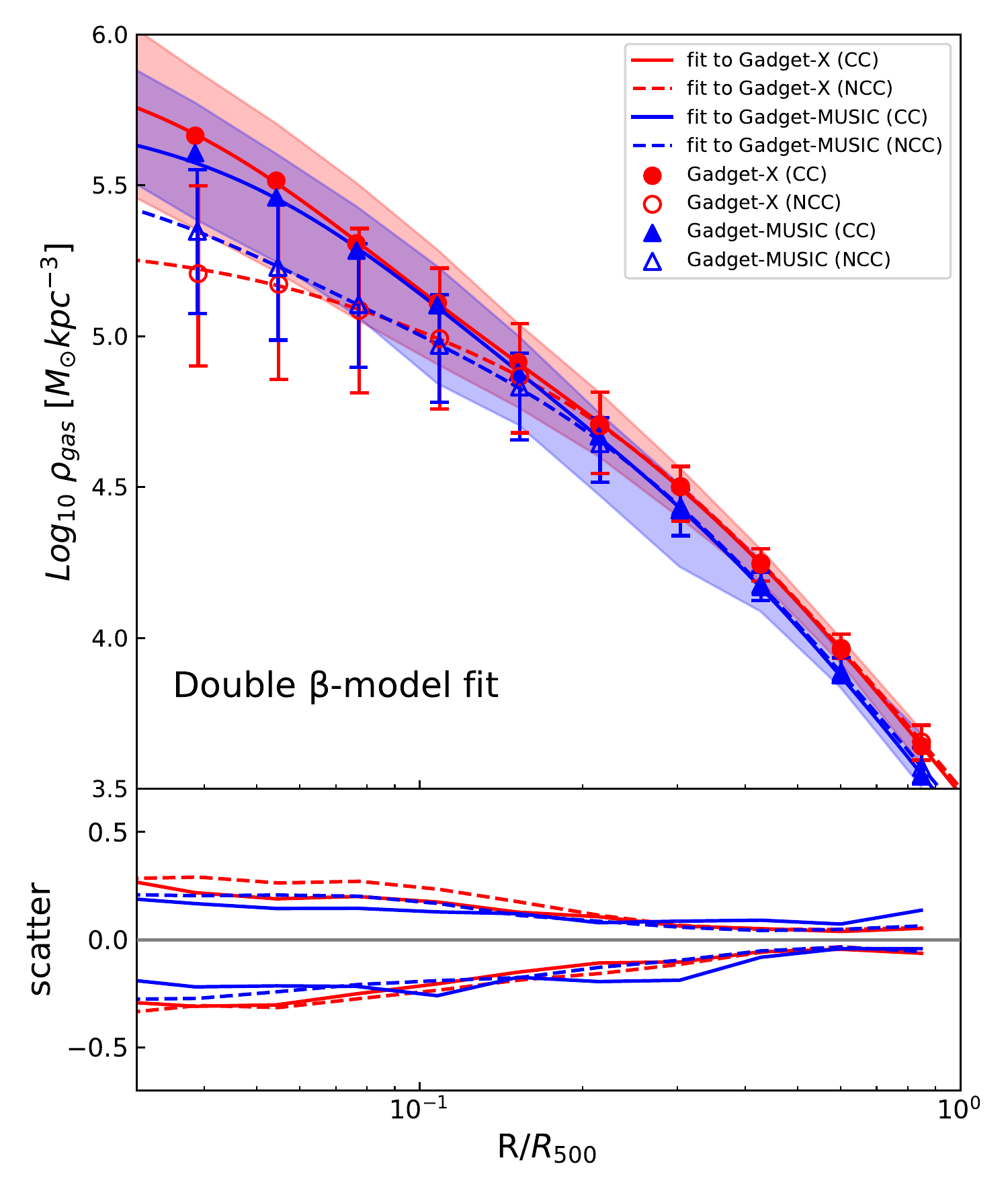}
    \caption{Similar to Fig.~\ref{fig:gas_density_state} with a separation of cool-core and non-cool-core clusters.}
    \label{fig:gas_density_cc}
\end{figure}

\begin{table*}
	\centering
	\caption{Similar to table~\ref{tab:gas_density_state} with fitting parameters for Fig. \ref{fig:gas_density_cc}.}
	\label{tab:gas_density_cc}
	\begin{tabular}{c c c c c c c}
	    \hline
        \multicolumn{2}{c}{fitting parameters}  & $\rho_1[\Msun/kpc^3]$ & $R_1[\rm R_{500}]$ & $\rho_2[\Msun/kpc^3]$ & $R_2[\rm R_{500}]$ & $\beta$ \\
		\hline
		model & core type \\
		\hdashline
		\multirow{2}{*}{ \gadgetx\ } & CC & $6.131 \times 10^4$ & 0.276 & $8.181 \times 10^5$ & 0.043 & 0.796  \\
		& NCC & $9.330 \times 10^4$ & 0.222 & $1.072 \times 10^5$ & 0.066 & 0.754 \\
		\hdashline
		\multirow{2}{*}{ \gadgetmusic\ } & CC & $5.056 \times 10^5$ & 0.060 & $4.369 \times 10^4$ & 0.288 & 0.813  \\
	    & NCC & $8.848 \times 10^4$ & 0.207 & $2.897 \times 10^5$ & 0.040 & 0.751  \\
		\hline
		fitting function & \multicolumn{6}{c}{$\rho(R) = \sum_{i=1}^2 \rho_i \left[1 + \left(\frac{R}{R_i}\right)^2\right]^{-3\beta/2}$}\\ 
		\hline
	\end{tabular}
\end{table*}

Gas density profiles are separated according to the cluster dynamical state in Fig.~\ref{fig:gas_density_state} and according to the CC/NCC dichotomy in Fig.~\ref{fig:gas_density_cc}. Since there is very little scatter in the gas density profiles at outer radii, we do not expect to see much difference at these radii. There is a clear separation in the cluster centre region -- un-relaxed and NCC clusters tend to have a lower density profile compared to relaxed and CC clusters. Thus only the inner gas density profile is influenced by the cluster dynamical state; the different trends of CC and NCC objects are in agreement with \citet{Ghirardini2019}. 

\subsection{The effects on gas temperature profile}

\begin{figure}
    \centering
    \includegraphics[scale = 0.5]{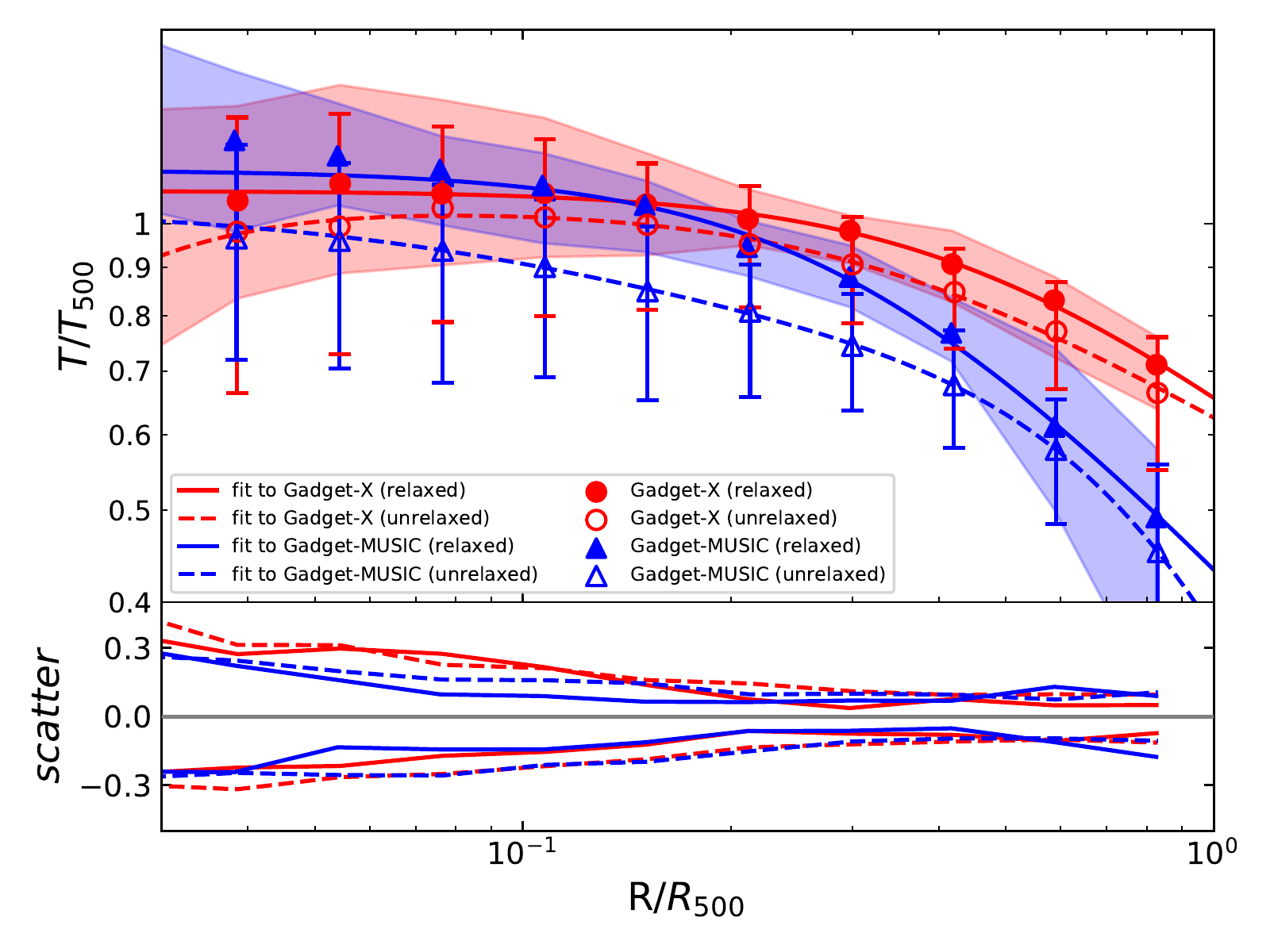}
    \caption{Similar to Fig.~\ref{fig:gas_density_state} but for the gas temperature profile of relaxed clusters and un-relaxed clusters. The fitting results shown by solid and dashed lines use the function from \citet{Ghirardini2019}.}
    \label{fig:gas_temp_state}
\end{figure}

\begin{table*}
	\centering
	\caption{The fitting function and parameters for Fig. \ref{fig:gas_temp_state}. The fitting formula is from \citet{Ghirardini2019}.}
	\label{tab:gas_temp_state}
	\begin{tabular}{c c c c c c c c}
		\hline
		\multicolumn{2}{c}{fitting parameters} & $T_0\ [T_{500}]$ & $T_{\rm min}\ [T_{500}]$ & $r_{\rm cool}\ [R_{500}]$  & $a_{\rm cool}$ & $r_t\ [R_{500}]$ & c\\
		\hline
		model & dynamical state \\
		\hdashline
		\multirow{2}{*}{ \gadgetx\ } & relaxed & 0.502 & 1.083 & 3.363 & 15.771 & 0.461 & 0.576 \\
		& un-relaxed & 1.041 & 0.619 & 0.021 & 2.965 & 0.330 & 0.440 \\
		\hdashline
		\multirow{2}{*}{ \gadgetmusic\ } & relaxed & 1.139 & 0.502 & 0.003 & 28.153 & 0.299 & 0.774 \\
	    & un-relaxed & 0.747 & 1.040 & 0.124 & 1.464 & 1.243 & 2.847 \\

		\hline
		fitting function & \multicolumn{7}{c}{$\frac{T(x)}{T_{500}} = T_0 \frac{\frac{T_{\rm min}}{T_0} + \left(\frac{x}{r_{\rm cool}}\right)^{a_{\rm cool}}}{1 + \left(\frac{x}{r_{\rm cool}}\right)^{a_{\rm cool}}} \frac{1}{\left(1 + \left( \frac{x}{r_t}\right)^2\right)^{\frac{c}{2}}}$} \\
		\hline
	\end{tabular}
\end{table*}

\begin{figure}
    \centering
    \includegraphics[scale = 0.5]{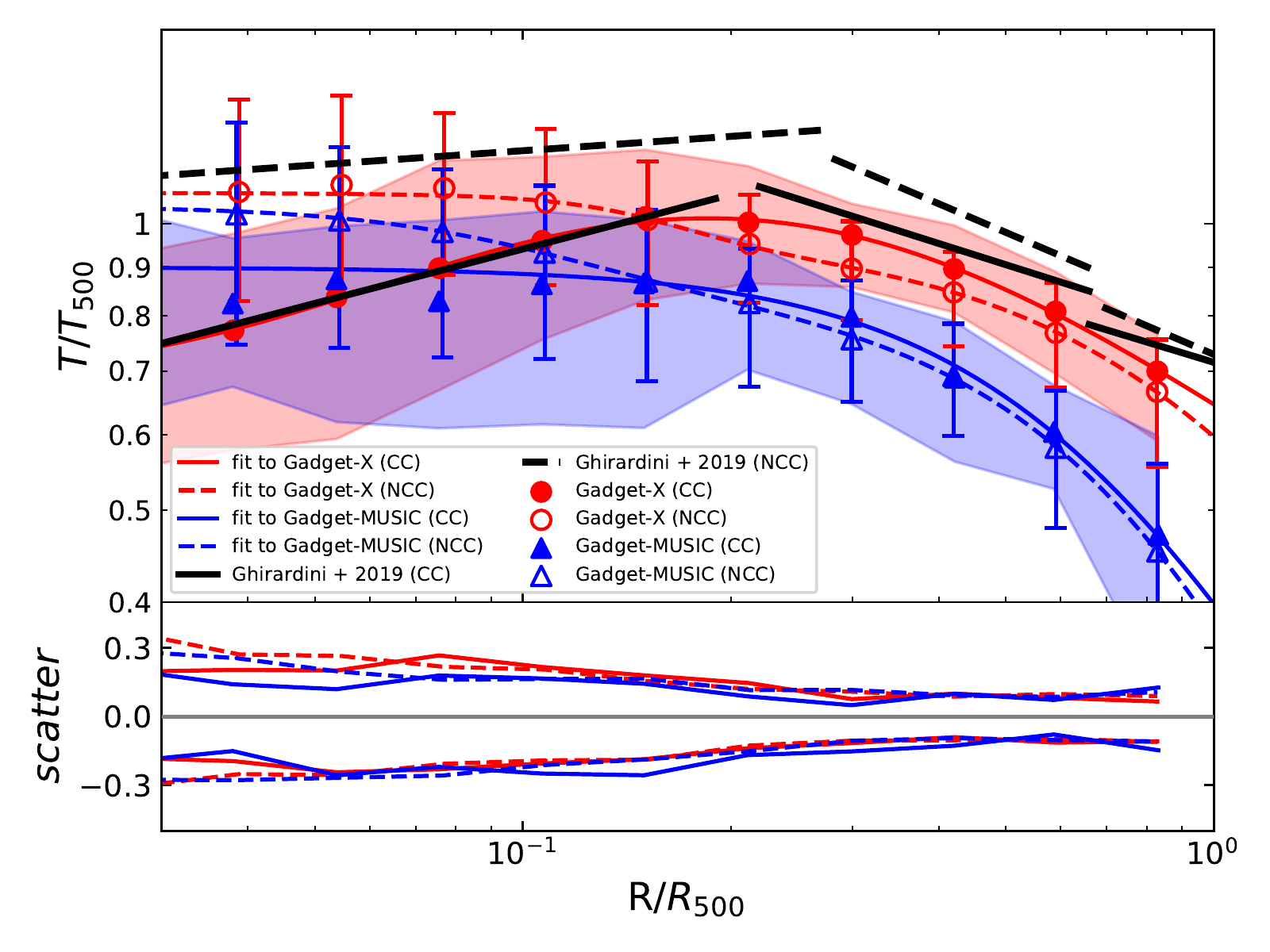}
    \caption{Similar to Fig.~\ref{fig:gas_density_state} but for the gas temperature profiles, separated according to cool-core clusters and non-cool-core clusters. The fitting results shown as solid and dashed lines use the equation from \citet{Ghirardini2019}.}
    \label{fig:gas_temp_cc}
\end{figure}

\begin{table*}
	\centering
	\caption{The fitting function and parameters for Fig. \ref{fig:gas_temp_cc}. The fitting formula is from \citet{Ghirardini2019}. The thick black solid and dashed lines are the CC and NCC clusters from \citet{Ghirardini2019}, repectively.}
	\label{tab:gas_temp_cc}
	\begin{tabular}{c c c c c c c c}
		\hline
		\multicolumn{2}{c}{fitting parameters} & $T_0\ [T_{500}]$ & $T_{\rm min}\ [T_{500}]$ & $r_{\rm cool}\ [R_{500}]$  & $a_{\rm cool}$ & $r_t\ [R_{500}]$ & c\\
		\hline
		model & core type \\
		\hdashline
		\multirow{2}{*}{ \gadgetx\ } & CC & 1.184 & 0.670 & 0.080 & 1.797 & 0.320 & 0.506 \\
		& NCC & 0.952 & 1.078 & 0.166 & 5.182 & 0.832 & 1.044 \\
		\hdashline
		\multirow{2}{*}{ \gadgetmusic\ } & CC & 0.900 & 0.485 & 0.021 & 31.268 & 0.641 & 1.321 \\
	    & NCC & 0.835 & 1.043 & 0.112 & 2.704 & 0.851 & 1.844 \\
		\hline
		fitting function & \multicolumn{7}{c}{$\frac{T(x)}{T_{500}} = T_0 \frac{\frac{T_{\rm min}}{T_0} + \left(\frac{x}{r_{\rm cool}}\right)^{a_{\rm cool}}}{1 + \left(\frac{x}{r_{\rm cool}}\right)^{a_{\rm cool}}} \frac{1}{\left(1 + \left( \frac{x}{r_t}\right)^2\right)^{\frac{c}{2}}}$} \\
		\hline
	\end{tabular}
\end{table*}

As with the gas density profiles, we also find that the dependence of the cluster dynamical state and CC/NCC dichotomy on gas temperature is much stronger in the cluster centre than at outer radii.

In both simulations, the dynamical relaxed clusters have a slightly higher temperature than these un-relaxed clusters, which tends to be more obviously in the centre than the outer radii. The scatter seems a little lower for the dynamical relaxed clusters than these un-relaxed clusters.

It is not surprising to see that the NCC clusters tend to have a higher temperature in the cluster centre than the CC clusters. However, unlike the gas density profile, the un-relaxed clusters tend to have a lower temperature. This could be understood as the gas in the un-relaxed clusters still being in the process of shock heating. As shown in Fig.~\ref{fig:gas_temp_cc}, the CC clusters in \gadgetx\ are in perfect agreement with \cite{Ghirardini2019}. The NCC clusters in \gadgetx\ seem to be lower than the result from \cite{Ghirardini2019} and have an opposite trend at outer radii, i.e. a lightly lower temperature rather than a higher temperature than the CC clusters which is presented in the observation result. Again, although \gadgetmusic\ give a similar result, it is systematically lower than the observations at outer radii which is caused by the normalisation.

\subsection{The effects on gas metallicity profile}
\begin{figure}
    \centering
    \includegraphics[scale = 0.5]{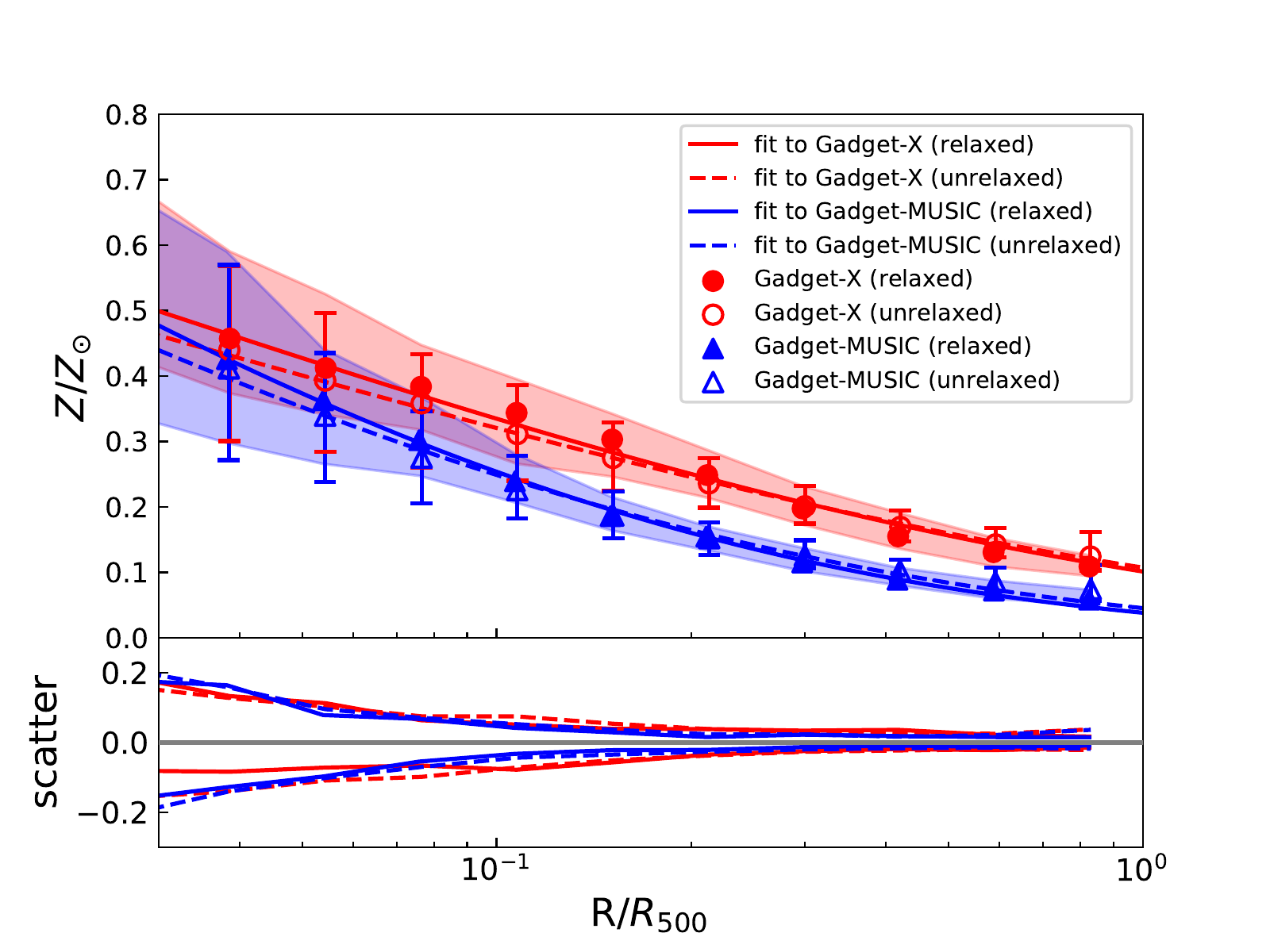}
    \caption{Similar to Fig.~\ref{fig:gas_density_state} but for the gas metallicity profile of relaxed clusters and un-relaxed clusters. The fitting results shown as solid and dashed lines use the S$\rm{\Acute{e}}$rsic profile (\citet{Sersic1963}).}
    \label{fig:gas_metal_state}
\end{figure}

\begin{table*}
	\centering
	\caption{The fitting function and parameters for Fig. \ref{fig:gas_metal_state}.}
	\label{tab:gas_metal_state}
	\begin{tabular}{c c c c c c}
		\hline
		fitting parameters & \multicolumn{2}{c}{\gadgetx} & & \multicolumn{2}{c}{\gadgetmusic}\\
		\hline
		dynamical state & relaxed & un-relaxed & & relaxed & un-relaxed\\
		\hline
		$Z_0[Z_{\odot}]$ & 0.060 & 0.146 & & 0.200 & 0.092\\
		$R_0[R_{500}]$ & 2.010 & 0.591 & & 0.146 & 0.448\\
		b & 3.254 & 2.192 & & 2.669 & 3.183\\
		\hline
		fitting function & \multicolumn{5}{c}{$Z(R)\ =\ Z_0e^{-b[\left(\frac{R}{R_0}\right)^{\frac{1}{4}} - 1]} $} \\
		\hline
	\end{tabular}
\end{table*}

\begin{figure}
    \centering
    \includegraphics[scale = 0.5]{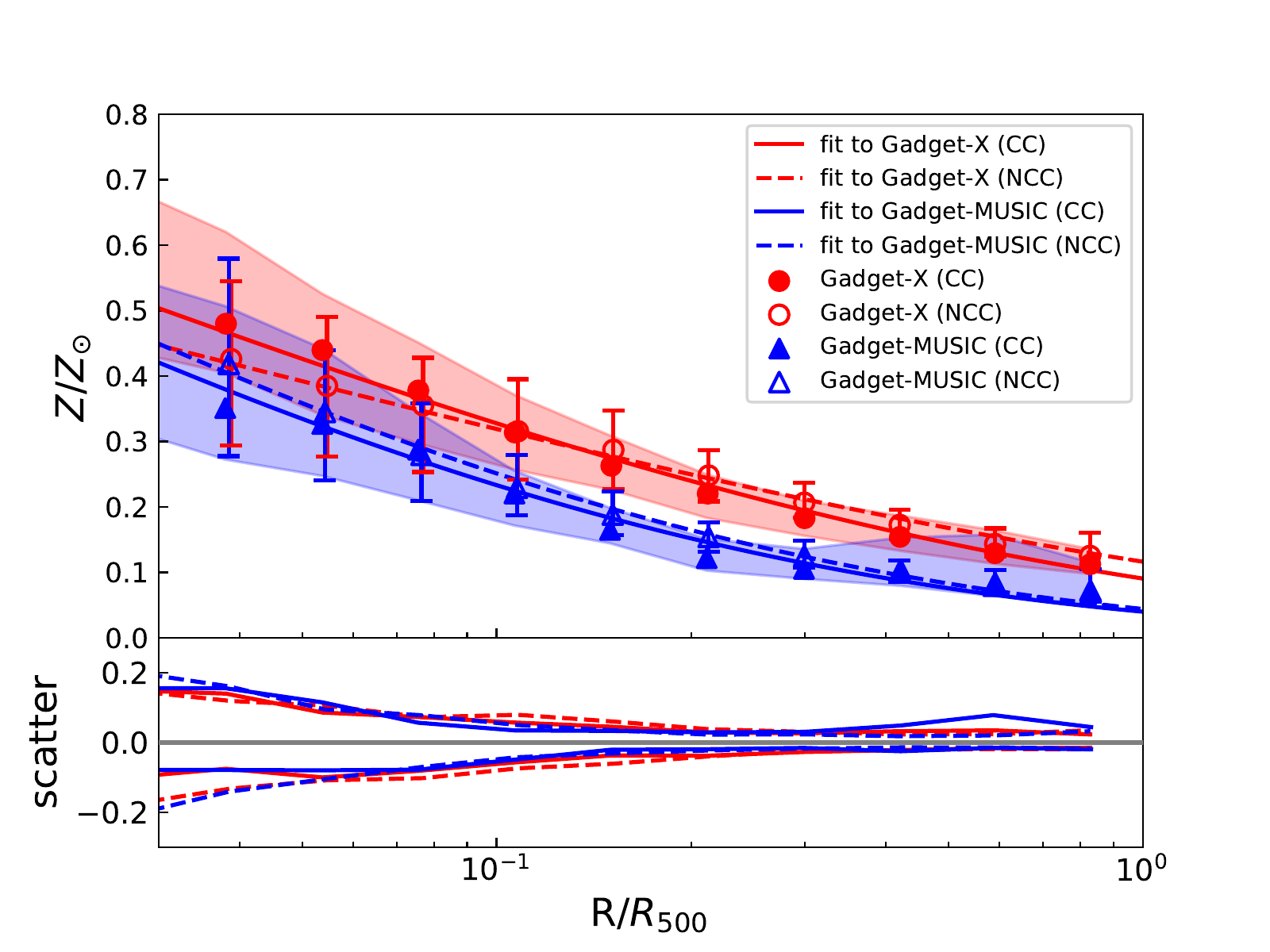}
    \caption{Similar to Fig.~\ref{fig:gas_density_state} but for the gas metallicity profile of CC and NCC clusters. The fitting results shown as solid and dashed lines use the S$\rm{\Acute{e}}$rsic profile (\citet{Sersic1963}).}
    \label{fig:gas_metal_cc}
\end{figure}

\begin{table}
	\centering
	\caption{The fitting function and parameters for Fig. \ref{fig:gas_metal_cc}.}
	\label{tab:gas_metal_cc}
	\begin{tabular}{c c c c c c}
		\hline
		fitting parameters & \multicolumn{2}{c}{\gadgetx} & & \multicolumn{2}{c}{\gadgetmusic}\\
		\hline
		core type & CC & NCC & & CC & NCC\\
		\hline
		$Z_0[Z_{\odot}]$ & 0.195 & 0.130 & & 0.171 & 0.091\\
		$R_0[R_{500}]$ & 0.297 & 0.816 & & 0.168 & 0.447\\
		b & 2.170 & 2.192 & & 2.581 & 3.253\\
		\hline
		fitting function & \multicolumn{5}{c}{$Z(R)\ =\ Z_0e^{-b[\left(\frac{R}{R_0}\right)^{\frac{1}{4}} - 1]} $} \\
		\hline
	\end{tabular}
\end{table}

As we expected, there is also not much difference between the metallicity profiles at the cluster outer radii between relaxed and un-relaxed clusters or CC and NCC clusters. While, in the cluster centre region, the differences are also very weak compared to the gas density or temperature profiles. In agreement with \citet{Lovisari2019}, relaxed clusters tend to have a higher metallicity in the centre for both \gadgetx\ and \gadgetmusic. However, given the large error bar, we would like to conclude that gas metallicity is likely to be less affected by the cluster dynamical state or CC/NCC dichotomy.

\section{The T500 difference} \label{app:2}

\begin{figure}
    \centering
    \includegraphics[scale=0.5]{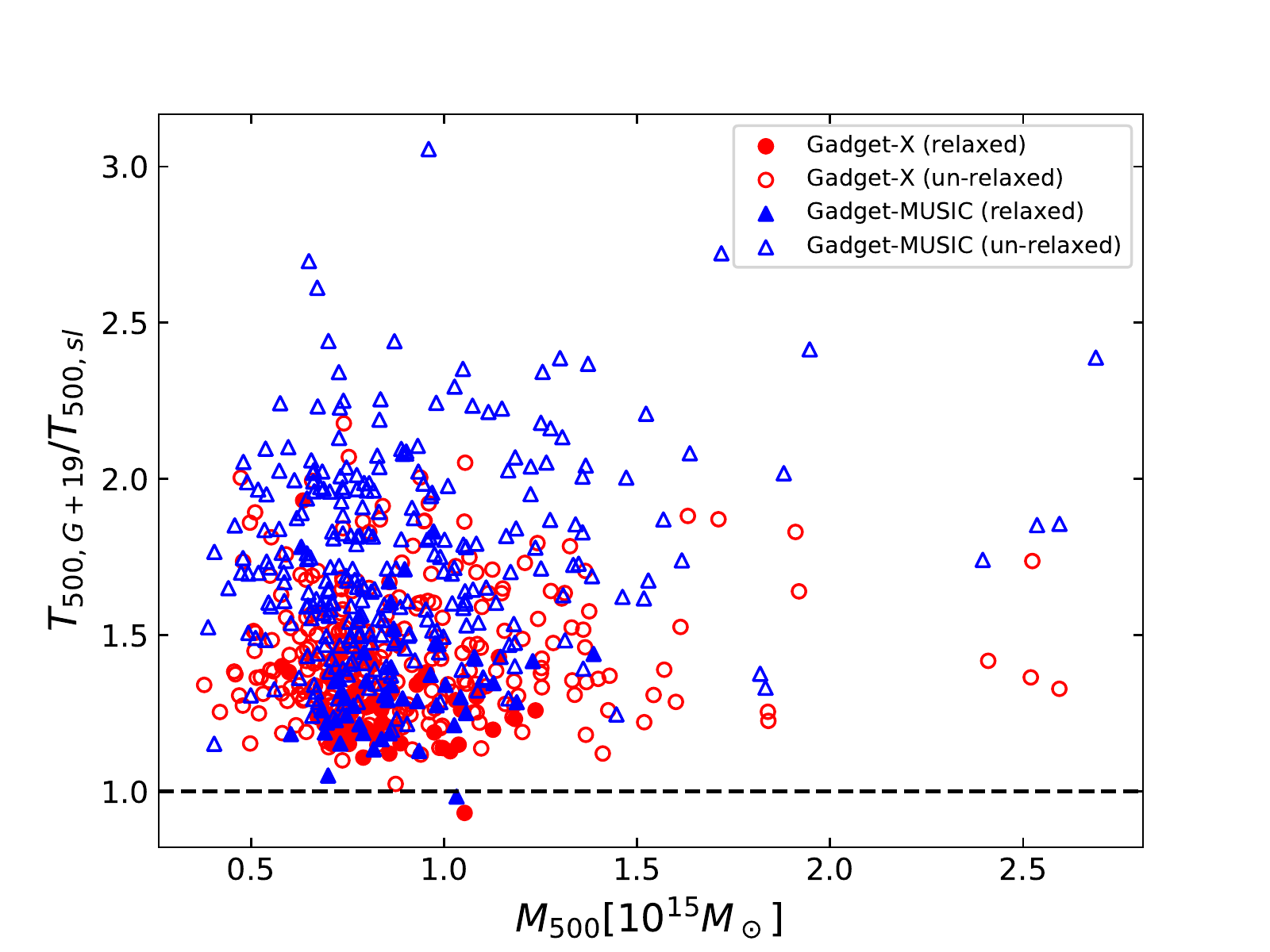}
    \caption{The ratio between $T_{500, G+19}$ and $T_{500, sl}$ as a function of halo mass $M_{500}$. Each symbol represents a cluster from \gadgetx (red circle) and \gadgetmusic (blue triangle) with filled for relaxed clusters and opened for un-relaxed clusters. }
    \label{fig:T500}
\end{figure}

In Fig.~\ref{fig:T500}, we show the ratio of cluster temperature $T_{500}$ between two different definitions which are illustrated in Subsection~\ref{subsec:gas temp}, respectively. The $T_{500, G+19}$ temperature is based on cluster mass $M_{500}$ (see equation \ref{T500G} for details), while $T_{500, sl}$ is calculated by spectroscopic weighted temperature using gas particles within a sphere of radius $R_{500}$. It is clear that $T_{500, G+19}$ is generally higher ($\sim 1.5$ times) than $T_{500, sl}$ with \gadgetmusic\ tends to have a higher difference and a larger scatter than \gadgetx. We do not see a clear mass dependence. By separating the clusters into relaxed and un-relaxed, it is clear that these clusters with large difference are basically un-relaxed clusters, which tend to give much larger biased halo mass estimated from the hydrostatic equilibrium assumption.


\bsp	
\label{lastpage}
\end{document}